\documentclass[sigconf,screen]{acmart}

\usepackage{multirow}
\usepackage{colortbl}
\usepackage{xcolor}
\usepackage{bbding}
\usepackage{appendix}
\usepackage{hyperref}
\usepackage{graphicx}
\usepackage{subcaption}

\AtBeginDocument{%
  }

\setcopyright{acmlicensed}
\copyrightyear{2025}
\acmYear{2025}
\acmDOI{XXXXXXX.XXXXXXX}
\acmConference[Conference acronym 'MM]{Make sure to enter the correct
  conference title from your rights confirmation email}{October 27--31, 2025}{Dublin, Ireland}
\acmISBN{978-1-4503-XXXX-X/25/10}

\acmSubmissionID{2118}

\begin{document}

\title{CRISP-SAM2 : SAM2 with Cross-Modal Interaction and Semantic Prompting for Multi-Organ Segmentation}

\author{Xinlei Yu}
\affiliation{%
  \institution{Hangzhou Dianzi University}
  \city{Hangzhou}
  \country{China}
}
\email{xinleiyu88@gmail.com}

\author{Changmiao Wang}
\authornote{Corresponding authors.}
\affiliation{
  \institution{Shenzhen Research Institute of Big Data}
  \city{Shenzhen}
  \country{China}
}
\email{cmwangalbert@gmail.com}

\author{Hui Jin}
\affiliation{
  \institution{Hangzhou Dianzi University}
  \city{Hangzhou}
  \country{China}
}
\email{hui1303101041@gmail.com}

\author{Ahmed Elazab}
\affiliation{
  \institution{Shenzhen University}
  \city{Shenzhen}
  \country{China}
}
\email{ahmedelazab@szu.edu.cn}

\author{Gangyong Jia}
\affiliation{
  \institution{Hangzhou Dianzi University}
  \city{Hangzhou}
  \country{China}
}
\email{gangyong@hdu.edu.cn}

\author{Xiang Wan}
\affiliation{
  \institution{Shenzhen Research Institute of Big Data}
  \city{Shenzhen}
  \country{China}
}
\email{wanxiang@sribd.com}

\author{Changqing Zou}
\affiliation{
  \institution{Zhejiang University}
  \city{Hangzhou}
  \country{China}
}
\email{changqing.zou@zju.edu.cn}

\author{Ruiquan Ge}
\authornotemark[1]
\affiliation{
  \institution{Hangzhou Dianzi University}
  \city{Hangzhou}
  \country{China}
}
\email{gespring@hdu.edu.cn}



\renewcommand{\shortauthors}{Xinlei Yu et al.}

\begin{abstract}
Multi-organ medical segmentation is a crucial component of medical image processing, essential for doctors to make accurate diagnoses and develop effective treatment plans. Despite significant progress in this field, current multi-organ segmentation models often suffer from inaccurate details, dependence on geometric prompts and loss of spatial information. Addressing these challenges, we introduce a novel model named CRISP-SAM2 with \textbf{CR}oss-modal \textbf{I}nteraction and \textbf{S}emantic \textbf{P}rompting based on \textbf{SAM2}. This model represents a promising approach to multi-organ medical segmentation guided by textual descriptions of organs. Our method begins by converting visual and textual inputs into cross-modal contextualized semantics using a progressive cross-attention interaction mechanism. These semantics are then injected into the image encoder to enhance the detailed understanding of visual information. To eliminate reliance on geometric prompts, we use a semantic prompting strategy, replacing the original prompt encoder to sharpen the perception of challenging targets. In addition, a similarity-sorting self-updating strategy for memory and a mask-refining process is applied to further adapt to medical imaging and enhance localized details. Comparative experiments conducted on seven public datasets indicate that CRISP-SAM2 outperforms existing models. Extensive analysis also demonstrates the effectiveness of our method, thereby confirming its superior performance, especially in addressing the limitations mentioned earlier. Our code is available at: \href{https://github.com/YU-deep/CRISP_SAM2.git}{https://github.com/YU-deep/CRISP\_SAM2.git}.
\end{abstract}

\ccsdesc[500]{Computing methodologies~Image segmentation}



\keywords{Multi-Organ Segmentation, Cross-Modal Semantic Interaction, Non-Geometric Prompting, Segment Anything Model}
\maketitle

\section{Introduction}
Efficient and accurate segmentation is crucial in modern medical and clinical applications, underpinning disease diagnosis, surgical planning, and outcome prediction~\cite{ma2024segment,cheng2024recent}. Multi-organ segmentation, in particular, has been a significant area of interest due to its clinical significance and technical challenges. Traditionally, this task has been performed manually by skilled radiologists or internists, which requires meticulous case-by-case annotation and verification. This manual process is not only time-consuming and labor-intensive but also prone to errors due to the high level of expertise needed. To address these challenges, researchers have developed relatively simple and lightweight networks~\cite{gibson2018automatic,wang2019abdominal,zhou2019prior,li2024magic}, which enable automatic segmentation but with limited accuracy. Recently, the Segment Anything Model (SAM)~\cite{kirillov2023segment} has demonstrated impressive capabilities in segmenting diverse objects with the help of prompts such as points, bounding boxes, and coarse masks. Its derivative, SAM2~\cite{ravi2024sam}, retains these capabilities while extending segmentation from image-level to video-level. Both SAM and SAM2 are trained on large-scale datasets, providing them with powerful zero-shot segmentation abilities in visual tasks.

\begin{figure}[t]
    \centering
    \begin{subfigure}[b]{0.48\textwidth}
        \centering
        \includegraphics[width=\textwidth]{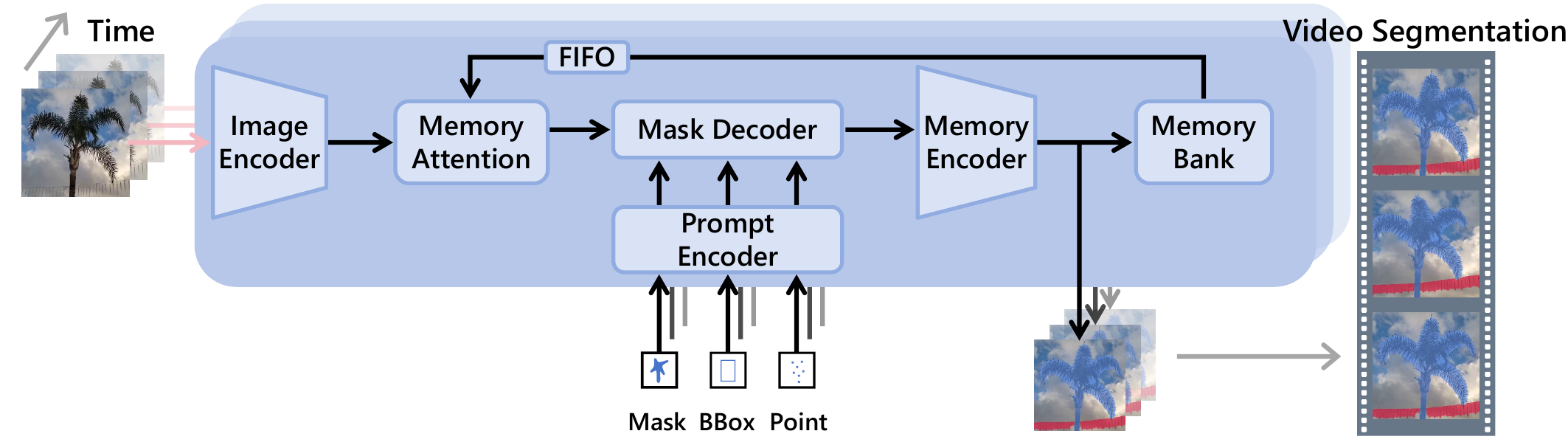}
        \caption{Segment Anything Model 2 (SAM2)}
    \end{subfigure}
    
    \begin{subfigure}[b]{0.48\textwidth}
        \centering
        \includegraphics[width=\textwidth]{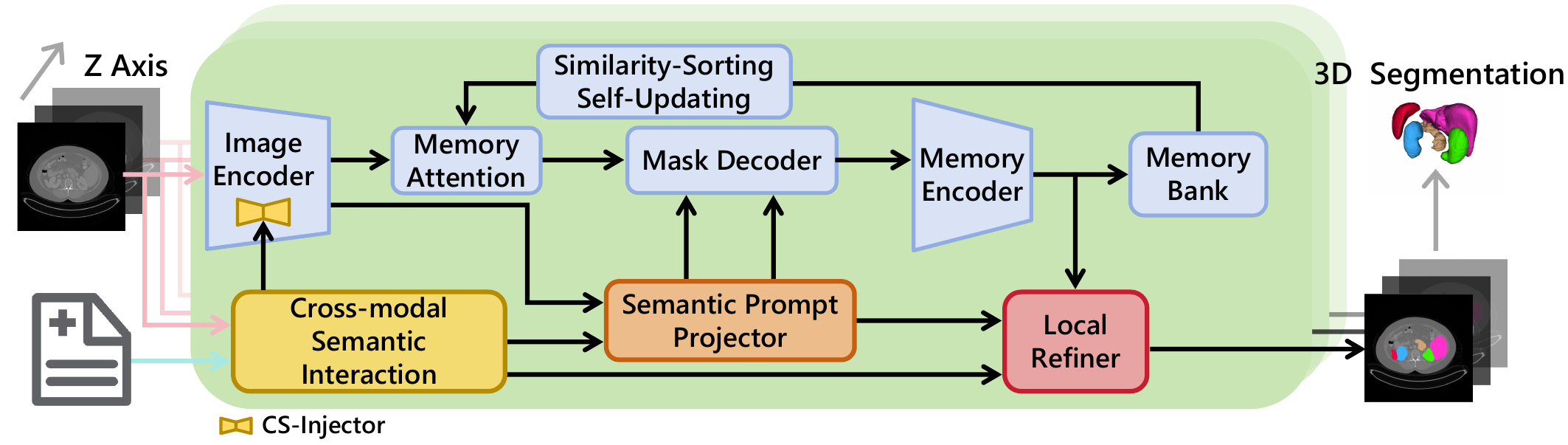}
        \caption{CRISP-SAM2 (Ours)}
    \end{subfigure}
    
    \begin{subfigure}[b]{0.48\textwidth}
        \centering
        \includegraphics[width=\textwidth]{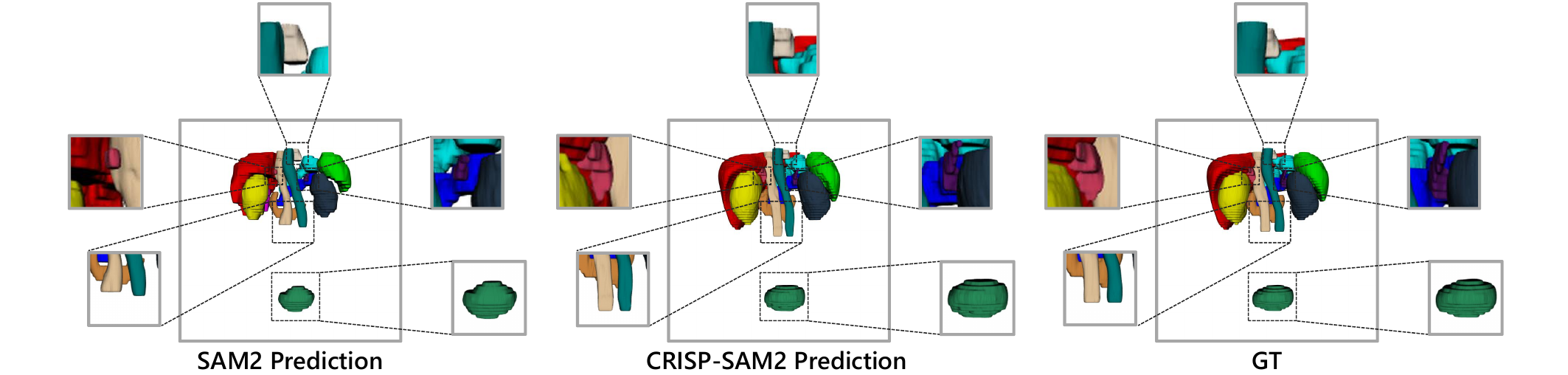}
        \caption{The Predicted Masks of SAM2 vs. CRISP-SAM2}
    \end{subfigure}
    \caption{Comparison between (a) SAM2~\cite{ravi2024sam} and (b) our CRISP-SAM2 designed for text-guided 3D multi-organ segmentation. Sub-figure (c) shows the predicted masks of SAM2 and our proposed CRISP-SAM2 against the ground truth.}
    \label{sam_com}
\end{figure}


While SAM, SAM2, and their variants have achieved significant success, there remains room for improvement, particularly in medical multi-organ segmentation. One of the primary challenges is related to their pre-training process~\cite{xin2024parameter}. Although these models are trained on extensive datasets featuring natural objects, the scarcity of medical data within these sets hinders the ability of the model to learn specific features of medical imaging effectively. As a result, the gap between the natural and medical domains remains unbridged. To address this, some researchers~\cite{ma2024segment,wang2023sam} have adapted SAM for medical segmentation by training it on large-scale medical datasets. Other models~\cite{wei2024medsam,guo2024towards,zhu2024medical,huang2024p2sam} have modified the original structures or added additional components to better accommodate the unique characteristics of medical imaging, thereby improving performance. Recently, innovative models~\cite{huang2025cat,jiang2024zept,du2025segvol} have begun to explore the integration of textual information to assist in segmentation. The rich contextual information inherent in natural language offers great potential for enhancing segmentation performance, pointing to promising directions for future research.

Current methods based on SAM or SAM2 face three main limitations: 1) They struggle with inaccurate local details and boundaries, sometimes including unwanted adjacent regions or excluding irregularly shaped targets from the predicted masks. 2) These methods rely heavily on explicit prompts, which can lead to inaccurate segmentation without geometric prompts, particularly for small or thin targets, thereby reducing their flexibility. 3) There is a loss of spatial information as they cannot directly segment 3D images. To overcome these challenges, we develop CRISP-SAM2, which leverages complementary textual information to guide visual segmentation, segmenting details and boundaries precisely and eliminating the need for geometric prompts.


To address the limitations mentioned earlier, we first introduce a two-level progressive cross-attention structure to generate cross-modal semantics. These semantics are integrated into the image encoder as contextualized information to enhance the understanding of visual features in detail. By converting these cross-modal semantics and image features into sparse and dense prompt embeddings, we can eliminate the dependency on geometric prompts. Besides, this semantic prompting method, which is guided by textual descriptions of organs, allows for capturing localized details and complex boundaries with improved accuracy. In addition to the original mask, we reuse the mask decoder on an additional learnable tokens and generate a refined mask, which is combined with the cross-modal semantics to further refine segmentation details. Recognizing the unique characteristics of 3D medical imaging and the ability of SAM2 to segment continuous imaging frames, we replace its original first-in-first-out (FIFO) strategy with a similarity-sorting self-updating strategy to better leverage spatial information. Our proposed model, CRISP-SAM2, based on SAM as shown in Figure~\ref{sam_com}, outperforming both purely visual models~\cite{ma2024segment,wei2024medsam,zhu2024medical,guo2024towards} and text-assisted models~\cite{li2023lvit,huang2025cat,du2025segvol,jiang2024zept} on seven public datasets. Comprehensive ablation studies and visualizations confirm the effectiveness and advantages of our designs. The key contributions of this work include:
\begin{itemize}
\item We propose CRISP-SAM2, a novel model specifically designed for accurate 3D multi-organ segmentation that leverages textual information of targets for guidance.
\item We develop a progressive cross-modal semantic interaction mechanism to integrate contextualized semantics, enhancing the understanding of cross-modal features.
\item We devise a novel prompting approach that uses cross-modal semantics for prompting and replaces the geometric prompts, refining the segmentation of challenging targets.
\item We conduct experiments with state-of-the-art (SOTA) models on seven public datasets, demonstrating the effectiveness of CRISP-SAM2, especially for details and boundaries.
\end{itemize}

\section{Related Works}
\subsection{Text-Guided Learning in SAMs and SAM2s}
In specific tasks, SAM and SAM2 may struggle to accurately segment target masks due to a lack of domain-specific training data. A potential solution to this issue involves integrating relevant textual features using an adapter or projector to guide the segmentation process. For instance, DIT~\cite{huang2024deep}, TP-DRSeg~\cite{li2024tp}, and LGA~\cite{hu2024lga} enhance SAM by injecting task-related textual features after each transformer block within the image encoder. In contrast, other approaches such as AdaptiveSAM~\cite{paranjape2024adaptivesam} introduces adapters in the mask decoder. Currently, the field of cross-modal interaction has seen significant advancements, particularly with the success of visual language models. Thus, the fusion of textual and visual embeddings is a logical step to create richer cross-modal features for guiding segmentation. RefSAM~\cite{Li2023RefSAMEA} employs multilayer perceptron (MLP) networks for this fusion, while SAP-SAM~\cite{wang2024fine} and Prompt-RIS~\cite{shang2024prompt} utilize cross-modal alignment and prompting modules. Similarly, EVF-SAM~\cite{zhang2024evf} uses a multi-modal encoder to produce fused embeddings. To effectively bridge the gap between textual and visual modalities, models like SEEM~\cite{zou2024segment} and Semantic-SAM~\cite{li2025segment} map texts, images, and prompts into a unified representation space, achieving semantic awareness. Moreover, Contrastive Language-Image Pre-Training (CLIP)~\cite{radford2021learning}, which excels in aligning and understanding semantics between texts and images, is employed in various segmentation tasks, such as open-vocabulary segmentation~\cite{yuan2025open,pan2025tokenize,wang2024samclip}, and other related fields~\cite{yu2024scnet,li2025clipsam}, offering valuable semantic insights.

\subsection{Prompt-Based Learning in SAMs and SAM2s}
To achieve precise segmentation using SAM and SAM2, it is essential to carefully design geometric prompts, namely points, bounding boxes, and coarse masks. In real-world applications, however, these auxiliary prompts may not always be readily available. Although SAM~\cite{kirillov2023segment} designs the use of text prompts, a model incorporating this feature is not released, and SAM2~\cite{ma2024segment} does not mention it. To address this, several models have been developed to generate geometric prompts or direct prompt embeddings. For instance, Grounding DINO~\cite{liu2025grounding} can detect objects based on textual input and is employed by various works~\cite{ren2024grounded,gowda2025cc} as a bounding box generator, which can also be combined with other prompts~\cite{zhang2023text2seg,cheng2023segment}. Besides, the range of generated prompts extends beyond bounding boxes to include points, coarse masks, and their various combinations. For example, GenSAM~\cite{hu2024relax} employs specialized CLIP and cross-modal reasoning to create points and bounding box prompts. SAM4MLLM~\cite{chen2025sam4mllm} and ZePT~\cite{jiang2024zept} utilize large language models to generate point prompts, while a cross-modal decoder is used in Prompt-RIS~\cite{shang2024prompt} to produce coarse masks. Models like SegVol~\cite{du2025segvol} and CAT~\cite{huang2025cat} combine textual semantic prompts with other types of prompts. Since the prompt encoder transforms prompts into sparse (points, bounding boxes) and dense (coarse masks) prompt embeddings, another approach is to directly generate these embeddings. For instance, RSPrompter~\cite{chen2024rsprompter} introduces a query- and anchor-based sparse embedding prompter. Similarly, SurgicalSAM~\cite{yue2024surgicalsam} and APSeg~\cite{he2024apseg} aim to produce both sparse and dense embeddings simultaneously by leveraging class prototype embeddings and the features of support-query pairs, respectively.


\section{Methodology}
The process of our proposed CRISP-SAM2 is divided into three distinct phases: Semantic Extraction, Prompt Generation, and Mask Prediction. In the Semantic Extraction phase, we introduce a Cross-Modal Semantic Interaction module to derive Cross-modal Semantic Features (CS-Features). These features are integrated into the image encoder of SAM2 as semantic guidance and serve as inputs for the subsequent phases. During the Prompt Generation phase, the Semantic Prompt Projector utilizes both the CS-Features and the four-scale features from the image encoder to produce both Sparse and Dense Prompt Embeddings. In the final Mask Prediction phase, we concatenate a learnable Cross-modal Output Tokens (CS-Output Tokens) with the original output and prompt tokens, which will be fed into the mask decoder together. Then, we use the Updated CS-Output Tokens as the input of the Local Refiner to correct the details of the predicted masks. Furthermore, we devise a similarity-sorting self-updating strategy tailored for 3D medical imaging segmentation to replace the original FIFO strategy.


\begin{figure*}[t]
    \centering
    \includegraphics[width=0.989\linewidth]{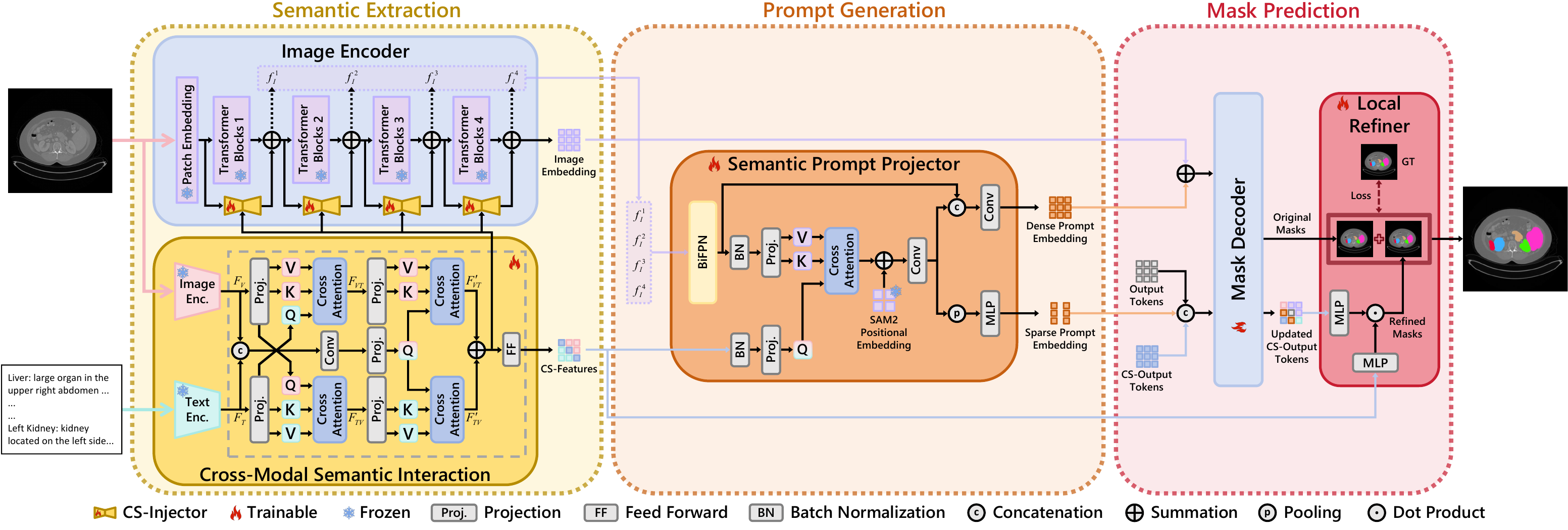}
    \caption{Overall structure of our CRISP-SAM2, which omits the memory attention, memory encoder, and memory bank components lfor clarity. CRISP-SAM2 produces accurate masks of 3D multi-organ segmentation under the guidance of textual information. A two-level progressive cross-modal interaction mechanism is adopted to extract contextualized semantics. Then, the semantics will be injected into image features, generate prompt embedding and further refine masks respectively, promoting superior segmentation prediction with precise local details and boundaries. Here, "Enc." represents "Encoder".}
    \label{overview}
\end{figure*}

\subsection{Preliminaries}
\label{preliminary}
We provide a brief overview of SAM2~\cite{ma2024segment}, which serves as the foundation of our CRISP-SAM2. It consists of six key components: (a) Image encoder: This utilizes a Hiera~\cite{ryali2023hiera} image encoder with a multi-scale hierarchical architecture, producing image embeddings through four-scale transformer blocks. (b) Prompt encoder: It encodes positional information of targets from geometric prompts into sparse and dense prompt embedding. (c) Mask decoder: This employs stacked two-way transformer blocks for mask prediction, leveraging image embeddings, prompt information, and output tokens. (d) Memory attention: Stacked transformer blocks that condition current features based on past ones. (e) Memory encoder: A lightweight convolution structure designed to generate memory. (f) Memory bank: Maintains a FIFO queue of memories to store past features. The first three components primarily focus  on mask prediction, which closely aligns with the architecture of SAM~\cite {kirillov2023segment} and modifies the image encoder and mask decoder. The latter three components enable it to segment video frames effectively. SAM2 has a three-stage training strategy for zero-shot generalization and is trained on an extensive dataset, which includes 50.9K videos with a total duration exceeding 196 hours. However, it is worth noting that training it from scratch is resource-intensive, requiring approximately 256$\times$108 A100 GPU hours. 


\subsection{Semantic Extraction}
\label{semantic_exrtaction_sec}
\subsubsection*{\textbf{Cross-Modal Semantic Interaction.}} To deepen the understanding of medical multi-organ segmentation, we introduce a module designed to integrate textual and visual inputs, generating complementary cross-modal semantics. This process is illustrated in the lower left part of Figure~\ref{overview}. These semantically enriched features, referred to as CS-Features $\mathit{\mathbf{F}_{CS}}$, are injected into successive transformer blocks across four scales and are utilized by both the Semantic Prompt Projector and Local Refiner.

Consider a pair of medical image $\mathbf{V}\in\mathbb{R}^{D_v\times H_v \times W_v}$ and its corresponding text description $\mathbf{T}\in\mathbb{R}^{L}$, where $D_v$, $H_v$, $W_v$ denote the depth, height, and width of the visual input, and $\mathit{L}$ is the length of the text input. For simplification, $\mathbf{V}$ is reduced to $\mathbf{V}^{\prime}\in\mathbb{R}^{H_v \times W_v}$, as all $D_v$ image slices share the same textual description. Operations related to image depth are handled by three memory components, which are detailed in Section~\ref{remaining_structure}. The preprocessing of texts involves appending $[SOS]$ and $[EOS]$ markers, followed by projecting it into a text embedding sequence $\mathbf{T}^{\prime}\in\mathbb{R}^{D_t \times \left(L+2\right)}$, following the CLIP method~\cite{radford2021learning}. The processed $\mathbf{V}^{\prime}$ and $\mathbf{T}^{\prime}$ are input into the image and text encoders in the Cross-Modal Semantic Interaction module respectively, producing visual and linguistic features $\mathit{\mathbf{F}_V}$ and $\mathit{\mathbf{F}_T}$.

Upon acquiring these features from separate modalities, we employ a progressive fusion structure with two levels of semantic interaction to derive the contextualized CS-Features from the paired images and texts. In the first interaction level, a multi-head cross-attention operation is performed between $\mathit{\mathbf{F}_V}$ and $\mathit{\mathbf{F}_T}$, preliminarily merging semantics from each modality. This is represented as:
\begin{equation}
  \begin{aligned}
  \mathit{\mathbf{F}_{VT}} = CrossAttn\left(\mathit{\mathbf{F}_T},\mathit{\mathbf{F}_V}\right),\\
    \mathit{\mathbf{F}_{TV}} = CrossAttn\left(\mathit{\mathbf{F}_V},\mathit{\mathbf{F}_T}\right), 
  \end{aligned}
  \label{cross_attention1}
\end{equation}
where the former parameter of $CrossAttn\left(\cdot,\cdot\right)$ is projected into the query for the multi-head cross-attention mechanism, and the latter from the other modality is projected into the key and the value. In order to capture localized information, we employ a depth-wise convolution projection following ~\cite{guo2022cmt} to replace direct linear projection, which is detailed in Appendix~\ref{methodology_appendix}. This initial interaction introduces features from one modality to the other, establishing preliminary dependencies between them.

To further extract complex cross-modal relationships and improve contextual representations, a second-level multi-head cross-attention layer is added. Initially, $\mathit{\mathbf{F}_V}$ and $\mathit{\mathbf{F}_T}$ are concatenated, preserving more efficacious information than a direct summation could, which reduces the loss of modal-specific features. A convolution layer is then applied to enhance consistent feature representation and eliminate redundancy from irrelevant features, producing the output:
\begin{equation}
    \mathit{\mathbf{F}_{C}} = Conv\left(Concat\left(\mathit{\mathbf{F}_V},\mathit{\mathbf{F}_T}\right)\right),
\end{equation}
where $Concat\left(\cdot,\cdot\right)$ denotes concatenation and $Conv\left(\cdot\right)$ represents convolution, using a kernel size of $3\times3$. $\mathit{\mathbf{F}_C}$ serves as the query for cross-attention with $\mathit{\mathbf{F}_{TV}}$ and $\mathit{\mathbf{F}_{VT}}$. In this interaction level, $\mathit{\mathbf{F}_C}$, which is enriched with consistent semantics from both visual and linguistic modalities, acts as a query to uncover potential complex correlations within initially fused cross-modal features $\mathit{\mathbf{F}_{TV}}$ and $\mathit{\mathbf{F}_{VT}}$ from previous level. This process is outlined as:
\begin{equation}
  \begin{aligned}
    \mathit{\mathbf{F}_{TV}^{\prime}} = CrossAttn\left(\mathit{\mathbf{F}_C},\mathit{\mathbf{F}_{TV}}\right), \\
    \mathit{\mathbf{F}_{VT}^{\prime}} = CrossAttn\left(\mathit{\mathbf{F}_C},\mathit{\mathbf{F}_{VT}}\right).
  \end{aligned}
  \label{cross_attention2}
\end{equation}

Finally, the outputs of the second-level attentions are summed, followed by a feed-forward (FF) layer, yielding the final CS-Features $\mathit{\mathbf{F}_{CS}}$ of this module. Notably, we inject these features into the image encoder prior to the FF layer to preserve original cross-modal semantics and relevant contexts, enhancing the semantic understanding of visual features given the robust feature extraction capabilities of transformer blocks. This progressive two-level cross-attention structure effectively extracts cross-modal semantics and relationships between visual and linguistic features. Additional experiments validating this two-level cross-modal semantic interaction strategy are presented in Section~\ref{ablation_sec}, comparing with other structures.

\subsubsection*{\textbf{Cross-Modal Semantic Injection.}} Recognizing that $\mathit{\mathbf{F}_{CS}}$ are unseen to the model, we employ a Cross-Modal Semantic Injector, i.e. CS-Injector, to integrate this complementary feature into the pre-trained image encoder of SAM2~\cite{ma2024segment}. This approach largely follows the architecture of SAM2-Adapter~\cite{chen2023sam,chen2024sam2}. Specifically, we frozen the pre-trained weights of the Hiera image encoder to preserve its strong visual representations. In addition, a simple and efficient CS-Injector is designed to inject semantics, consisting of just two light-weight MLPs. The dimensions of the second MLP correspond to the four hierarchical transformer blocks within the image encoder. It is worth noting that we additionally sum the output of the first MLP with the original input of visual features as the input of the second MLP. Ultimately, the injected image encoder produces the Image Embedding $\mathit{\mathbf{E}_{I}}$.

\subsection{Prompt Generation}
\label{prompt_generation_sec}
Despite the impressive segmentation capabilities of SAM~\cite {kirillov2023segment} and SAM2~\cite{ma2024segment}, geometric prompts are still essential for accurate segmentation, especially in challenging scenarios where accurate points, bounding boxes, and masks are absent~\cite{xie2024pa}. Medical images often pose such challenges due to their fuzzy and irregularly shaped boundaries, increasing the need for precise prompts. However, obtaining these prompts requires interactive labeling by experienced doctors, which significantly raises labor and time costs and limits broader applications of the models in medical contexts.

To address this limitation, we introduce an efficient Semantic Prompt Projector. This innovation automatically generates Sparse and Dense Prompt Embeddings without additional demanding of auxiliary geometric inputs. The Semantic Prompt Projector utilizes the CS-Features and outputs from the four transformer blocks within the image encoder and applies positional embedding to guarantee that the Sparse and Dense Prompt Embeddings share the same position information with the Image Embedding.

As illustrated in the middle of Figure~\ref{overview}, we initially derive visual features at different scales from four outputs of each transformer block. These features, denoted as $\{\mathit{f_{I}^{\mathrm{1}}}, \mathit{f_{I}^{\mathrm{2}}}, \mathit{f_{I}^{\mathrm{3}}}, \mathit{f_{I}^{\mathrm{4}}}\}$, are processed through a BiFPN~\cite{tan2020efficientdet} structure to create $\mathit{\mathbf{F}_I^{\prime}}$, synthesizing multiscale information. Next, batch normalization (BN)~\cite{ioffe2015batch} is applied to both visual features and CS-Features to balance them, which is followed by a multi-head cross-attention to produce the Semantic Embedding: 
\begin{equation}
  \mathit{\mathbf{E}_{S}} = CrossAttn\left(LN\left(\mathit{\mathbf{F}_{CS}}\right),LN\left(\mathit{\mathbf{F}_I^{\prime}}\right)\right).
  \label{cross_attention3}
\end{equation}

Since the image encoder remains frozen during training and inference, the Position Embedding $\mathit{\mathbf{E}_P}$ can be computed and stored previously. By summing $\mathit{\mathbf{E}_P}$ with $\mathit{\mathbf{E}_{S}}$, we enhance the capture of positional information within the prompt embedding, which is subsequently processed through a convolution layer:
\begin{equation}
  \mathit{\mathbf{E}_{S}^{\prime}} = Conv\left( \mathit{\mathbf{E}_{S}} \oplus \mathit{\mathbf{E}_{P}}\right),
\end{equation}
where $\oplus$ denotes summation. In SAM2~\cite{ma2024segment}, sparse and dense prompt embeddings are derived from points/bounding boxes and coarse masks, providing geometric guidance for segmentation. The former provides crude location information, while the latter gives a more dense and comprehensive representation of the overall structure of segmentation targets. Emulating this design, we concatenate $\mathit{\mathbf{E}_{S}^{\prime}}$ with the visual feature $\mathit{\mathbf{F}_I^{\prime}}$ to enhance the visual representation, while using a pooling layer to sparsify representation. It should be noted that we use the ada-pooling~\cite{stergiou2022adapool} to avoid the loss of details with efficiency while being suitable for sparse coding. Then, the final Dense Prompt Embedding $\mathit{\mathbf{E}_{DP}}$ and Sparse Prompt Embedding $\mathit{\mathbf{E}_{SP}}$ are obtained through an extra convolution layer and MLP:
\begin{equation}
  \begin{aligned}
\mathit{\mathbf{E}_{DP}}=Conv\left(Concat\left(\mathit{\mathbf{F}_I},\mathit{\mathbf{E}_{S}^{\prime}}\right)\right),\\
\mathit{\mathbf{E}_{SP}}=MLP_{2}\left(Pool\left(\mathit{\mathbf{E}_{S}^{\prime}}\right)\right),
  \end{aligned}
\end{equation}
where $MLP_{2}\left(\cdot\right)$ refers to the two-layer MLP and $Pool\left(\cdot\right)$ denotes the ada-pooling~\cite{stergiou2022adapool} layer. These prompt embeddings, generated by the Semantic Prompt Projector, are then used by the mask decoder to aid in segmentation.

\subsection{Mask Prediction}
\label{mask_predicion}
\subsubsection*{\textbf{Mask Decoding.}} Partly inspired by the design of HQ-SAM~\cite{ke2024segment}, we introduce an additional learnable output tokens, termed CS-Output Tokens $\mathit{\mathbf{T}_{CS}}$, which matches the size of the original Output Tokens. As one input of the mask decoder, $\mathit{\mathbf{T}_{CS}}$ is concatenated with the original Output Tokens $\mathit{\mathbf{T}_{O}}$ and the Sparse Prompt Embedding $\mathit{\mathbf{E}_{SP}}$. Following the original design, the Image Embedding $\mathit{\mathbf{E}_{I}}$ is combined with the Dense Prompt Embedding $\mathit{\mathbf{E}_{DP}}$. Consequently, the Original Masks segmentation is expressed as:
\begin{equation}
\mathit{\mathbf{M}_{O}} = \mathit{D_{s}}\left(\mathit{\mathbf{E}_{I}}\oplus\mathit{\mathbf{E}_{DP},Concat\left(\mathit{\mathbf{T}_{O}},\mathit{\mathbf{E}_{SP}},\mathit{\mathbf{T}_{CS}}\right)}\right),
\end{equation}
where $\mathit{D_{s}}$ represents the mask decoder. Additionally, the mask decoder of SAM2~\cite{ma2024segment} incorporates stride 4 and 8 features from the image encoder. During this process, $\mathit{\mathbf{T}_{CS}}$ gains access to the global visual context, geometric prompt information, and cross-modal semantics, evolving into the Updated CS-Output Tokens $\mathit{\mathbf{T}_{CS}^{\prime}}$.

\subsubsection*{\textbf{Local Refining.}} With the Original Masks $\mathit{\mathbf{M}_{O}}$ obtained, we proceed to generate Refined Masks $\mathit{\mathbf{M}_{R}}$ for correction. This involves applying two three-layer MLP on the Updated CS-Output Tokens $\mathit{\mathbf{T}_{CS}^{\prime}}$ and CS-Features $\mathit{\mathbf{F}_{CS}}$ respectively, followed by a dot product layer to produce $\mathit{\mathbf{M}_{R}}$. The predicted logits of $\mathit{\mathbf{M}_{O}}$ and $\mathit{\mathbf{M}_{R}}$ are then summed to refine the prediction, resulting in the final segmentation mask $\mathit{\mathbf{M}_{F}}$. This process is represented by:
\begin{equation}
\mathit{\mathbf{M}_{F}} = \left(MLP_3\left(\mathit{\mathbf{F}_{CS}}\right)\odot MLP_3\left(\mathit{\mathbf{T}_{CS}^{\prime}}\right)\right) \circ \mathit{\mathbf{M}_O},
\end{equation}
where $MLP_3\left(\cdot\right)$ denotes a three-layer MLP, $\odot$ signifies the dot product operation, and $\circ$ indicates logits summation.

\subsection{Remaining Components}
\label{remaining_structure}
\subsubsection*{\textbf{Similarity-Sorting Self-Updating Strategy.}} Distinct from natural videos, FIFO is unsuitable for direct application to 3D medical imaging due to the specific characteristics of these images. For example, initial and final slices of abdominal scans often contain only background, with no organs present. Inputting these slices first and including these slices in the memory bank could adversely affect the segmentation accuracy of other slices. To address this, we implement a similarity-sorting self-updating strategy to ensure only high-quality slices are retained in the memory bank. We begin by calculating a similarity score for each slice in a 3D image, using the sum of cosine similarity between each slice and all others. The slices are then ranked based on their similarity score to establish a segmentation sequence that prioritizes slices to be segmented with higher similarity scores, thus providing more accurate guidance for subsequent slices. To update the memory bank, we first filter obtained embedding by the intersection over union (IoU) score from the mask decoder and then calculate the embedding similarity scores of the current frame and previous embeddings stored in the memory bank, then, replace the embedding with the lowest score. More realization details are in Appendix~\ref{methodology_appendix}.


\subsubsection*{\textbf{Loss Function.}} In line with SAM2~\cite{ma2024segment}, we employ focal and dice losses for both Original Masks $\mathit{\mathbf{M}_{O}}$ and Refined Masks $\mathit{\mathbf{M}_{R}}$ to supervise mask predictions. Additionally, a mean-absolute-error (MAE) loss is applied for IoU prediction, and cross-entropy (CE) loss is used for object detection. The overall loss function is organized as:
\begin{equation}
\mathcal{L} = \alpha\mathcal{L}_{seg}^{original} + (1 - \alpha)\mathcal{L}_{seg}^{refined} + \beta\mathcal{L}_{mae} + \gamma\mathcal{L}_{ce},
\end{equation}
where $\alpha, \beta$, and $\gamma$ are learnable weights, $\mathcal{L}_{mae}$ denotes the MAE loss, and $\mathcal{L}_{ce}$ represents the CE loss. The segmentation losses $\mathcal{L}_{seg}^{original}$ and $\mathcal{L}_{seg}^{refined}$ are calculated as: $\mathcal{L}_{seg} = \omega_{1}\mathcal{L}_{focal} + \omega_{2}\mathcal{L}_{dice}$, with a weight ratio of $\omega_{1}:\omega_{2} = 20:1$, which is consistent with the original setting of SAM2.

\begin{table*}[t]
\centering
\caption{Segmentation performance of our proposed method and other comparison models on seven datasets. The results are evaluated by the average DSC and NSD of all the segmented organs. $*$ means assisted with geometric prompts, and $\diamond$ denotes assisted with anatomical prompt (refer to~\cite{huang2025cat}). "Abdomen" refers to "AbdomenCT-1k", and the same in the following tables.}
\setlength{\tabcolsep}{0.9mm}{
\resizebox{0.843\linewidth}{!}{
\begin{tabular}{c|cc|cc|cc|cc|cc|cc|cc}
\toprule
 \multirow{2}{*}{Method} & \multicolumn{2}{c|}{MSD-Spleen} & \multicolumn{2}{c|}{Pancreas-CT} & \multicolumn{2}{c|}{LUNA16} & \multicolumn{2}{c|}{Abdomen} & \multicolumn{2}{c|}{WORD}  & \multicolumn{2}{c|}{FLARE22} & \multicolumn{2}{c}{AMOS22}  \\ \cline{2-3} \cline{4-5} \cline{6-7} \cline{8-9} \cline{10-11} \cline{12-13} \cline{14-15}
 &DSC&NSD&DSC&NSD&DSC&NSD&DSC&NSD&DSC&NSD&DSC&NSD&DSC&NSD\\ \midrule
SAM$^*$~\cite{kirillov2023segment}&73.82&70.58&65.08&60.41&56.74&51.44&64.63&60.47&63.26&60.29&64.62&60.77&62.27&58.81\\
SAM2$^*$~\cite{ravi2024sam}&83.60&79.09&79.92&75.33&63.38&60.60&83.94&79.11&76.54&74.48&78.19&76.95&74.73&72.55\\ \midrule
MedSAM$^*$~\cite{ma2024segment}&76.07&70.95&70.08&66.52&60.81&54.37&69.07&64.05&72.47&68.75&67.43&64.36&65.10&61.76\\
I-MedSAM$^*$~\cite{wei2024medsam}&81.37&78.70&75.14&73.84&60.63&58.99&74.71&69.82&67.60&63.35&74.34&71.70&70.08&66.39\\
MedSAM-2$^*$~\cite{zhu2024medical}&86.29&86.24&79.48&80.33&64.00&65.62&86.76&86.41&77.42&78.61&84.81&85.50&79.31&80.64\\ 
CT-SAM3D$^*$~\cite{guo2024towards}&91.83&93.56&84.57&85.65&72.61&76.97&90.05&\underline{94.55}&80.33&83.99&89.99&\underline{94.85}&85.74&88.54\\\midrule
LViT~\cite{li2023lvit}&79.40&78.98&76.54&77.34&70.56&74.78&75.95&76.97&73.66&74.03&75.14&72.35&72.60&69.12\\
CAT$^{\diamond}$~\cite{huang2025cat}&\underline{94.26}&\underline{95.82}&\underline{86.29}&88.13&\underline{77.31}&80.53&88.38&92.79&83.53&88.40&\textbf{90.42}&94.06&86.39&89.25\\
SegVol$^*$~\cite{du2025segvol}&92.79&93.83&81.10&83.66&75.20&77.94&87.70&91.35&81.65&85.78&82.68&82.83&80.15&80.67\\
ZePT~\cite{jiang2024zept}&94.03&95.75&85.71&\underline{89.30}&77.17&\underline{83.21}&\underline{91.34}&94.37&\underline{84.26}&\underline{88.96}&86.28&90.01&\textbf{87.03}&\underline{89.27}\\
\rowcolor{gray!20}
\textbf{CRISP-SAM2 (Ours)}&\textbf{95.33}&\textbf{97.26}&\textbf{87.22}&\textbf{90.65}&\textbf{78.58}&\textbf{85.15}&\textbf{92.28}&\textbf{96.40}&\textbf{85.47}&\textbf{90.51}&\underline{90.30}&\textbf{95.59}&\underline{86.88}&\textbf{90.74}  \\ \bottomrule
\end{tabular}}}
\label{comparison}
\end{table*}

\section{Experiments}
\subsection{Experimental Setups}
\subsubsection*{\textbf{Datasets.}} Our training and inference experiments are conducted on M3D-Seg dataset~\cite{bai2024m3d}, a cross-modal medical segmentation joint dataset. M3D-Seg includes a variety of classical medical sub-datasets, each annotated with visual segmentation targets and textual descriptions consisting of two to six sentences. We specifically focus on seven high-quality sub-datasets from M3D-Seg: MSD-Spleen~\cite{simpson2019large}, Pancreas-CT~\cite{clark2013cancer}, LUNA16~\cite{setio2017validation}, AbdomenCT-1k~\cite{ma2021abdomenct}, WORD~\cite{luo2022word}, FLARE22~\cite{ma2023unleashing}, and AMOS22~\cite{ji2022amos}. These sub-datasets have one to fourteen segmentation categories, encompassing eighteen abdominal and thoracic organs: liver (L), spleen (SP), stomach (ST), gallbladder (G), esophagus (E), pancreas (P), duodenum (D), aorta (A), bladder (B), inferior vena cava (IV), left kidney (LK), right kidney (RK), left adrenal gland (LA), right adrenal gland (RA), left femur (LF), right femur (RF), left lung (LL) and right lung (RL). We maintain the original data partition of M3D-Seg~\cite{bai2024m3d}, which has an 8:2 ratio between training and testing sets. To prevent data bias and misrepresentation, we exclude partial labels of organs, tumors, and nodules that appear in only one sub-dataset. Notably, the AbdomenCT-1k~\cite{ma2021abdomenct} dataset originally label both the left and right kidneys as kidneys, thus we differentiate these into LK and RK by connected components and coordinates. Furthermore, the postcava label in AMOS22~\cite{ji2022amos} and the inferior vena cava label in FLARE22~\cite{ma2023unleashing} dataset refer to the same organ actually, which we consistently denote as IV. For additional details and illustrations regarding the selected datasets, please refer to Appendix~\ref{appendix_datasets}.

\subsubsection*{\textbf{Implementation Details.}} To expand the datasets and balance class of samples, we employ extensive data augmentation techniques. During the training phase, CRISP-SAM2 model uses a two-stage strategy for a total of 400 epochs, incorporating a warm-up period. We use the AdamW optimizer~\cite{loshchilov2017decoupled} with an initial learning rate of \(1e^{-4}\) and a polynomial decay power of 0.9. All experiments are performed on 8 NVIDIA A100-SXM4-80G GPUs with a batch size set to 16. Additional details are provided in Appendix~\ref{implementation_appendix}.

\subsubsection*{\textbf{Evaluation Metrics.}} We use the Dice Similarity Coefficient (DSC, \%) and Normalized Surface Distance (NSD, \%) as evaluations, which are commonly used in medical segmentation tasks. Both metrics range from 0 to 1, with higher values indicating more accurate predictions. DSC emphasizes the overall overlap between predicted masks and ground truths, while NSD is more sensitive to the consistency of boundaries and localized details between them.


\begin{figure*}[t]
    \centering
    \includegraphics[width=0.883\linewidth]{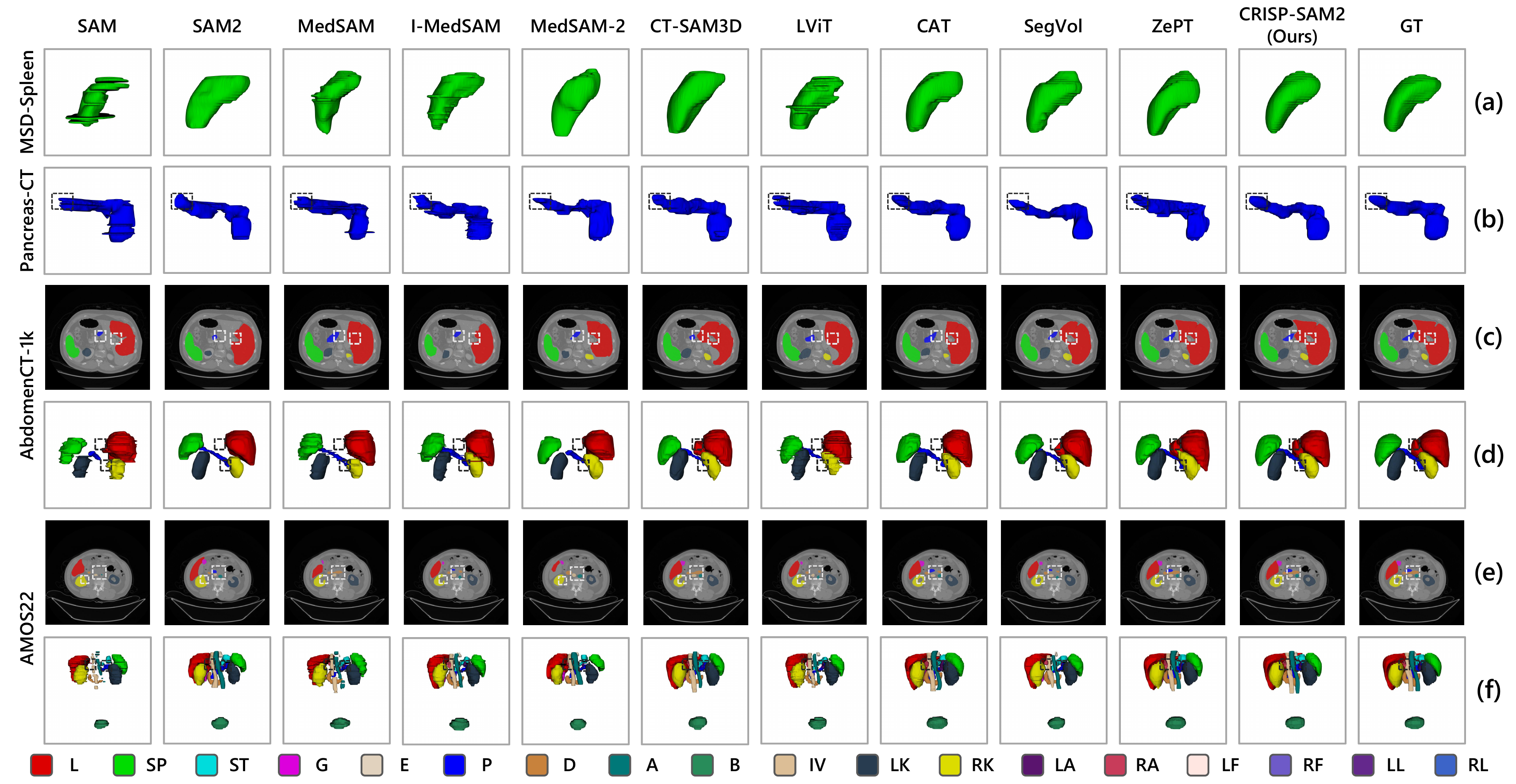}
   \caption{Visualization demonstrations of CRISP-SAM2 and other counterpart SOTA methods on selected datasets.  Rows (a), (b), (d), and (f) show the segmentation results of selected cases in 3D view, while rows (c) and (e) show the predicted masks in 2D imaging slices. The areas enclosed by the dashed black or white boxes showcase some predicted details among different models.}
    \label{visualization}
\end{figure*}

\subsection{Comparison Results}
\label{comparison_experiments}
To evaluate the performance of our CRISP-SAM2 model, we perform comparative experiments against other SOTA methods in the seven datasets, utilizing SAM~\cite{kirillov2023segment} and SAM2~\cite{ravi2024sam} models as baselines. Our comparisons include both purely visual segmentation approaches, namely MedSAM~\cite{ma2024segment}, I-MedSAM~\cite{wei2024medsam}, MedSAM-2~\cite{zhu2024medical}, and CT-SAM3D~\cite{guo2024towards}, and text-engaged segmentation methods, including LViT~\cite{li2023lvit}, CAT~\cite{huang2025cat}, SegVol~\cite{du2025segvol}, and ZePT~\cite{jiang2024zept}. For the comparison of prompt-required models, we utilize three points and a bounding box as prompts, which are randomly selected from the center of ground truth and a rectangle completely covers targets. More comparison results are shown in Appendix~\ref{main_experiment_appendix}.

\subsubsection*{\textbf{Quantitative Results.}} Table~\ref{comparison} presents the segmentation performance on the test sets of the seven datasets. Our model significantly outperforms the baselines, achieving at least a 21.5\% and 7.3\% improvement in the DSC, and at least a 26.6\% and 15.3\% increase in NSD compared to SAM and SAM2, respectively. Compared to other models, CRISP-SAM2 shows superior performance, with improvements in average DSC and NSD ranging from 0.93\% to 1.27\% and 1.26\% to 1.94\% across the MSD-Spleen~\cite{simpson2019large}, Pancreas-CT~\cite{clark2013cancer}, LUNA16~\cite{setio2017validation}, AbdomenCT-1k~\cite{ma2021abdomenct}, and WORD~\cite{luo2022word} datasets. On the FLARE22~\cite{ma2023unleashing} and AMOS22~\cite{ji2022amos} datasets, CRISP-SAM2 maintains a lead of 0.74\% to 1.47\% in NSD, highlighting its accuracy in segmenting local details, while being only marginally less than 0.12\% and 0.15\% behind the best in DSC.

Compared to unimodal medical segmentation methods~\cite{ma2024segment,wei2024medsam,zhu2024medical,guo2024towards}, CRISP-SAM2 achieves the highest DSC and NSD values across all datasets, with at least 2.99\% and 4.03\% enhancement in DSC and NSD evenly. Among all the text-guided methods~\cite{li2023lvit,huang2025cat,du2025segvol,jiang2024zept}, CRISP-SAM2 demonstrates the best overall performance, with an average improvement of at least 0.74\% in DSC and 1.62\% in NSD across the seven datasets. Specifically, on the AbdomenCT-1k~\cite{ma2021abdomenct} dataset, which is the largest of the seven, with 800 training samples and 200 test samples, CRISP-SAM2 outperforms the four relatively outstanding comparative methods, i.e. CT-SAM3D~\cite{guo2024towards}, CAT~\cite{huang2025cat}, SegVol~\cite{du2025segvol} and ZePT~\cite{jiang2024zept} with DSC and NSD improvements of 2.23\%/3.90\%/4.58\%/0.94\% and 1.85\%/3.61\%/5.05\%/2.03\%, respectively, demonstrating superior segmentation accuracy.

Furthermore, CRISP-SAM2, prompting by cross-modal semantics without reliance on geometric prompts, outperforms prompt-required methods, which showcases the effectiveness of the Semantic Prompt Projector module. We will discuss this part further in Section~\ref{ablation_sec}. It is also noteworthy that models integrating textual information, namely CAT~\cite{huang2025cat}, SegVol~\cite{du2025segvol}, ZePT~\cite{jiang2024zept}, and our CRISP-SAM2, generally perform better than uni-modal visual methods, particularly on datasets like LUNA16~\cite{setio2017validation} and WORD~\cite{luo2022word} where other methods underperform. This is because these text-guided models incorporate complementary information from textual inputs, which assists with segmentation. In summary, our CRISP-SAM2 could segment accurately with details and boundaries consistent with ground truths, exhibiting superior overall performance than other SOTA models, especially in challenging scenarios.

\subsubsection*{\textbf{Visualization Results.}} Figure~\ref{visualization} presents the visualization results of our CRISP-SAM2 model alongside other SOTA methods, providing additional evidence of the effectiveness and superiority of our method. The proposed CRISP-SAM2 excels particularly in dealing with local details and irregularly shaped boundaries. For instance, row (b) and the dashed boxes in rows (c) and (d) highlight how CRISP-SAM2 produces more precise boundaries for irregularly shaped organs. Additionally, rows (e) and (f) demonstrate CRISP-SAM2 superior ability to accurately segment localized details, especially for small and slender organs. These results intuitively demonstrate that CRISP-SAM2 addresses common challenges faced by existing methods, achieving superior segmentation performance.

\subsection{Ablation Studies}
\label{ablation_sec}
In our ablation study, we conduct several experiments to demonstrate the effectiveness of our proposed method and investigate aspects that influence its performance, including component ablation, an analysis of the Cross-Modal Semantic Interaction module, the effectiveness of the Semantic Prompt Projector, and qualitative results based on textual inputs. All experiments are conducted on the AbdomenCT-1k~\cite{ma2021abdomenct} and AMOS22~\cite{ji2022amos} datasets to ensure fair comparison, keeping other experimental settings constant.

\subsubsection*{\textbf{Components Ablations.}} Our method mainly adds four key improvements to the baseline SAM2: the Cross-Modal Semantic Interaction module, the Semantic Prompt Projector, the Local Refiner, and the similarity-sorting self-updating strategy. We select SAM2 as a baseline and design experiments with or without these components and combinations of components. First, we add the Cross-modal Semantic Interaction module to the baseline, whose output is indispensable for the Semantic Prompt Projector and the Local Refiner. We retain the CS-injector to integrate cross-modal semantics and directly output masks from the mask decoder. Next, we evaluate the performance improvements by incorporating the Semantic Prompt Projector, the Local Refiner, and both together. In addition, we assess the performance of the similarity-sorting self-updating strategy both separately and with other components.

As shown in Table~\ref{ablation}, the Cross-Modal Semantic Interaction module effectively fuses textual and visual features, generating complementary cross-modal semantics. This integration leads to improvements of 3.42\%-4.78\% in DSC and 5.21\%-6.08\% in NSD, underscoring the value of textual guidance in segmentation. Building on this, the inclusion of the Semantic Prompt Projector further enhances performance by more than 4\% in DSC and 5.5\% in NSD, aided by the generated Sparse and Dense Prompt Embeddings. Although the Local Refiner does not significantly improve DSC, it notably boosts NSD, indicating more precise segmentation of localized and small details. Compared to the FIFO strategy, our similarity-sorting self-updating strategy, tailored to the characteristics of medical images, resulted in an increase of over 0.4\% and 1.3\% in the two metrics. The model incorporating all four key components outperforms those with only individual components or other combinations, demonstrating that each component positively contributes and their integration is complementary.

\begin{table}[t]
\centering
\caption{Components ablation results on two selected datasets. In this table, SI, PP, LR, and SS denote Cross-Modal Semantic Interaction, Semantic Prompt Projector, Local Refiner, and similarity-sorting self-updating strategy respectively.}
\setlength{\tabcolsep}{0.9mm}{
\resizebox{0.775\linewidth}{!}{
\begin{tabular}{cccc|cc|cc}
\toprule
  \multicolumn{4}{c|}{Component} & \multicolumn{2}{c|}{Abdomen} & \multicolumn{2}{c}{AMOS22}  \\ \cline{1-4} \cline{5-6} \cline{7-8} 
 SI & PP & LR & SS & DSC & NSD & DSC & NSD\\ \midrule
&&&& 83.94 & 79.11 & 74.73 & 72.55 \\
 \CheckmarkBold &&&& 87.36 & 85.19 & 79.51 & 77.76\\
 \CheckmarkBold & \CheckmarkBold &&& 90.42 & 90.75 & 84.20 & 83.58\\
 \CheckmarkBold && \CheckmarkBold & & 86.78 & 90.34 & 81.06 & 82.96\\
 \CheckmarkBold & \CheckmarkBold & \CheckmarkBold & & 91.93 & 95.11 & 86.32 & 89.39  \\
 &&& \CheckmarkBold & 84.64 & 83.07 & 75.55 & 75.61\\
 \rowcolor{gray!20}
\CheckmarkBold & \CheckmarkBold & \CheckmarkBold & \CheckmarkBold & \textbf{92.28} & \textbf{96.40} & \textbf{86.88} & \textbf{90.74}  \\ \bottomrule
\end{tabular}}}
\label{ablation}
\end{table}

\begin{table}[t]
\centering
\caption{Analytical experiments between our two-level mechanism in the Cross-modal Semantic Interaction module and other interaction structure.}
\setlength{\tabcolsep}{0.9mm}{
\resizebox{0.845\linewidth}{!}{
\begin{tabular}{c|cc|cc}
\toprule
 \multirow{2}{*}{Interaction Method}  & \multicolumn{2}{c|}{Abdomen} & \multicolumn{2}{c}{AMOS22}  \\ \cline{2-3} \cline{4-5}
  & DSC & NSD & DSC & NSD\\ \midrule
$\Rightarrow$first & 89.51 & 92.77 & 85.10 & 88.39 \\
$\Rightarrow$second & 87.94 & 91.04 & 84.83 & 87.12 \\
$\Rightarrow$second$\Rightarrow$first & 91.59 & 95.43 & 86.21 & 89.56 \\
 \rowcolor{gray!20}
\textbf{$\Rightarrow$first$\Rightarrow$second} & \textbf{92.28} & \textbf{96.40} & \textbf{86.88} & \textbf{90.74}  \\ \bottomrule
\end{tabular}}}
\label{semantic_interaction}
\end{table}

\subsubsection*{\textbf{Analysis of Semantic Interaction.}} To effectively integrate visual and linguistic features, we develop a two-level progressive cross-attention semantic interaction strategy. To assess the impact of this approach, we conduct experiments analyzing the effects of using each level of cross-attention separately and in combinations. As shown in Table~\ref{semantic_interaction}, our interaction structure achieves the best performance, surpassing the next best structure, which reverses the order of the two levels of cross-attention, by more than 0.6\% in DSC and 0.9\% in NSD. The two-level cross-attention approach extracts richer and more profound cross-modal semantics compared to single-level structures, leading to improved segmentation of localized details. Benefiting from our proposed structure, the first level of cross-attention establishes an initial dependency between the two separate modalities, while the second level strengthens this dependency and effectively extracts cross-modal semantics.

\subsubsection*{\textbf{Effectiveness of Prompt Generation.}} Most segmentation models heavily rely on geometric prompts, i.e. points, bounding boxes, and coarse masks, to achieve accurate results. It significantly limits their wider applications, thus, we conduct further experiments to validate the effectiveness of our Semantic Prompt Projector. In these experiments, we replace the Semantic Prompt Projector with the original prompt encoder and compare various combinations of geometric prompts. Given that coarse masks are rarely available in real medical settings and using other models to generate them could introduce bias, we exclude mask prompts from our comparison. As shown in Table~\ref{prompt_generation}, the improvement brought by prompts is obvious, and our textual prompt could achieve superior performance compared to the use of point and bounding box prompts, which improves the DSC and NSD by over 2.3\% and 3.6\%, respectively. When compared to the model without any prompts, these improvements increase to over 5.0\% and 5.5\%, respectively.


\begin{table}[t]
\centering
\caption{Experimental results of the comparisons between geometric point prompts, bbox prompts, and our proposed textual prompting.}
\setlength{\tabcolsep}{0.9mm}{
\resizebox{0.8\linewidth}{!}{
\begin{tabular}{ccc|cc|cc}
\toprule
 \multicolumn{3}{c|}{Prompt}  & \multicolumn{2}{c|}{Abdomen} & \multicolumn{2}{c}{AMOS22}  \\ \cline{1-3} \cline{4-5} \cline{6-7} 
Text & Point & BBox & DSC & NSD & DSC & NSD\\ \midrule
 & & & 87.20 & 90.89 & 81.49 & 83.63 \\
 & \CheckmarkBold & &  88.99 & 91.33 & 83.50 & 84.76 \\
 & & \CheckmarkBold & 89.17 & 91.85 & 83.94 & 85.02 \\
 & \CheckmarkBold & \CheckmarkBold & 89.78 & 92.72 & 84.57 & 86.28 \\
 \rowcolor{gray!20}
 \textbf{\CheckmarkBold} & & & \textbf{92.28} & \textbf{96.40} & \textbf{86.88} & \textbf{90.74}  \\ \bottomrule
\end{tabular}}}
\label{prompt_generation}
\end{table}

\section{Conclusion}
In this study, we introduce CRISP-SAM2, an innovative model tailored for medical multi-organ segmentation that effectively integrates cross-modal interaction and semantic prompting. This approach utilizes additional textual information, specifically descriptions of target organs, to guide segmentation predictions. To generate cross-modal semantics, we develop a Cross-Modal Semantic Interaction module featuring a two-level cross-attention interaction mechanism. This mechanism progressively extracts contextualized semantics from distinct visual and textual modalities. Leveraging the obtained cross-modal semantics, the Semantic Prompt Projector produces Sparse and Dense Prompt Embeddings without relying on geometric prompts, while the Local Refiner and similarity-sorting self-updating strategy further enhance the precision of boundaries and local details in the predicted masks. Our model successfully addresses existing challenges in the field, namely inaccuracies in details and boundaries, reliance on geometric prompts, and compromised spatial information. Extensive experiments, including comparisons with other SOTA methods, component ablation studies, further analysis, and visualization results, highlight the superiority and efficacy of CRISP-SAM2, which obtains superior performance than other SOTA models on seven publicly available datasets.



\section*{Acknowledgments}
This work was supported by the Open Project Program of the State Key Laboratory of CAD\&CG (No. A2410), Zhejiang University; National Natural Science Foundation of China (No. 61702146, 62076084, U20A20386, U22A2033); Guangdong Basic and Applied Basic Research Foundation (No. 2025A1515011617); Guangxi Key R\&D Project (No. AB24010167).

\nocite{yu2025prnet}

\bibliographystyle{ACM-Reference-Format}
\bibliography{main}

\clearpage
\appendix
\section*{Appendix}
\section{Related Works}
\subsection{SAM and SAM2}
Deep learning has made impressive progress in several fields, such as computer vision~\cite{lu2023tf,li2025finecir,li2025semi,chen2025dyconfidmatch,li2024towards,wu2025prompt,song2025lvpnet,zeng2024residential,zhang2025marl,chen2024learning,lu2024mace,qian2024maskfactory,yao2025rethinking,chen2023class,xiao2025td,song2025bpclip,li2024deep,lu2024robust,yu2025prnet,li2024distinct,gao2024eraseanything,zhang2025gamed,cai2024msdet,chen2024sam,zeng2025automated,yao2024event,li2025encoder,song2024adtah}, natural language processing~\cite{bi2024visual,bi2025prism,chen2024bimcv,chen2023bridging,li2024towards1,xiao2025describe}, time series analysis~\cite{qiu2024tfb,wu2025k,qiu2025duet,wu2024catch,qiu2025tab,liu2025rethinking,qiu2025comprehensive}, multi-modal analysis~\cite{wu2025generative,zeng2024mitigating,guo2025each,gu2025mathcal,li2025theory,wu2025image,li2025set}, and multi-task learning~\cite{xin2024vmt,xin2024mmap,wu2024tai,wu2025multi}. However, there is still a lack of unified models in the field of computer vision, which is the reason why SAM and SAM2 are proposed.

The SAM framework~\cite{kirillov2023segment} is built around three primary components: an image encoder, a prompt encoder, and a mask decoder. Initially, the image to be segmented is processed through the image encoder to generate the image embedding. Subsequently, the mask decoder utilizes this image embedding, along with a prompt embedding derived from points, bounding boxes, and coarse masks via the prompt encoder, to predict the final masks. SAM2~\cite{ravi2024sam} retains the core structure but incorporates several enhancements, allowing it to segment videos as a series of consecutive frames. This capability is primarily facilitated by three additional components: memory attention, a memory encoder, and a memory bank. Memory attention conditions the current frame based on previous frames and prompts, linking features of each frame with those in the temporal dimension. The memory encoder generates memories by integrating the image embedding of an unconditioned frame. These memories are then stored in the memory bank in a FIFO sequence.

\begin{figure}[t]
    \centering
    \includegraphics[width=0.90\linewidth]{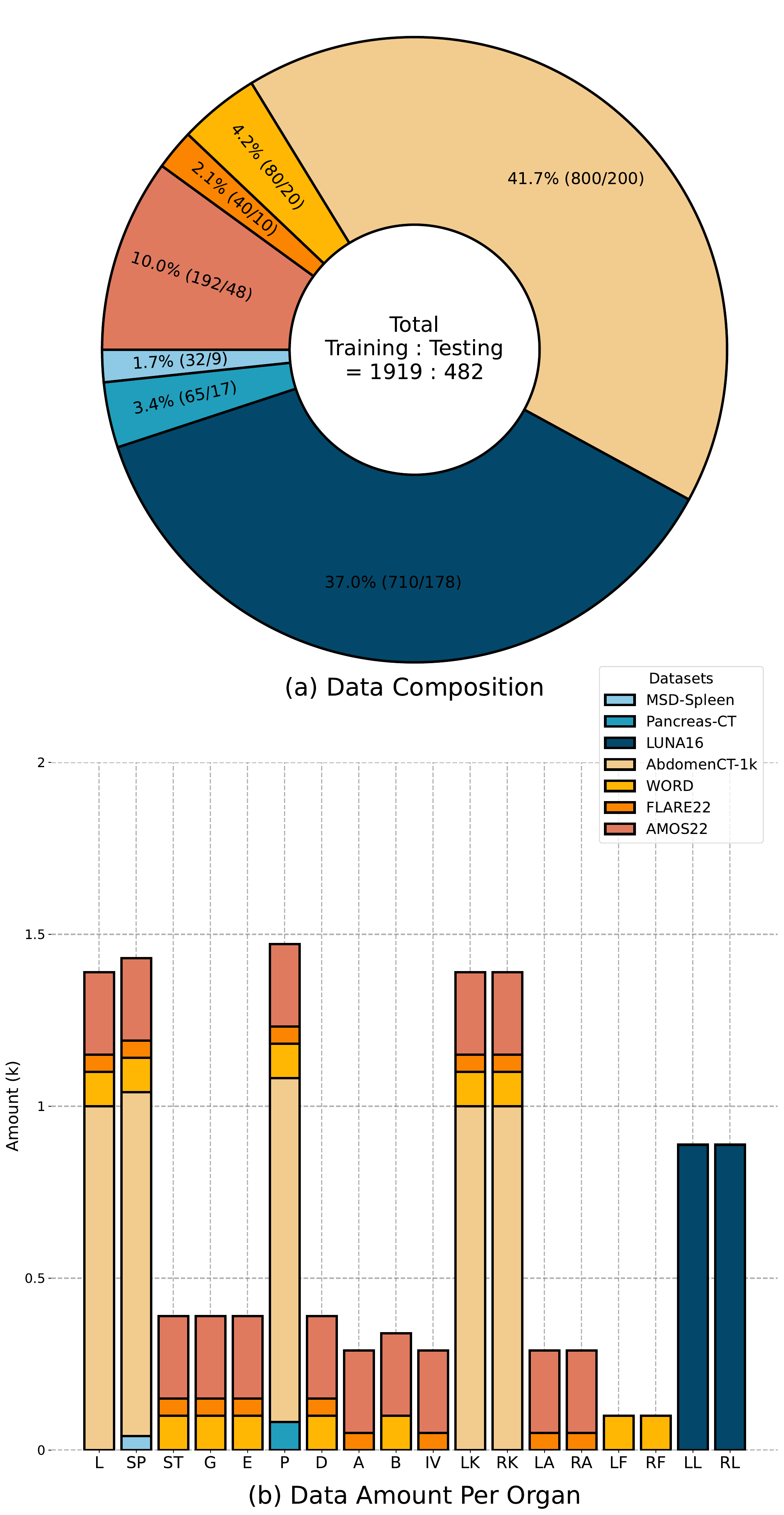}
   \caption{The upper (a) annular chart shows the composition of the training and testing samples, and the lower (b) bar chart illustrates the amount and composition of samples of each organ.}
    \label{dataset_png}
\end{figure}

\begin{table*}[t]
\centering
\caption{Details of seven selected datasets. Here, we use capital letters to represent each organ.}
\setlength{\tabcolsep}{0.9mm}{
\resizebox{0.80\linewidth}{!}{
\begin{tabular}{c|c|c|c|c|c|c|c|c|c|c|c|c|c|c|c|c|c|c|c|c}
\toprule
 \multirow{2}{*}{Dataset} & \multicolumn{18}{c|}{Organ} & \multirow{2}{*}{Train/Test} & \multirow{2}{*}{Category}  \\
 &  L & SP & ST & G & E & P & D  & A  & B & IV & LK & RK & LA & RA & LF & RF & LL & RL &  & \\ \midrule
  MSD-Spleen~\cite{simpson2019large}  & & \checkmark &&&&&&&&&&&&&&&& &  32/9 & 1\\
  Pancreas-CT~\cite{clark2013cancer}  &&&&&&\checkmark&&&&&&&&&&&&& 65/17 & 1\\
  LUNA16~\cite{setio2017validation}  &&&&&&&&&&&&&&&&&\checkmark&\checkmark& 710/178 & 2\\ 
  AbdomenCT-1k~\cite{ma2021abdomenct} &\checkmark&\checkmark&&&&\checkmark&&&&&\checkmark&\checkmark&&&&&&& 800/200 & 5\\
   WORD~\cite{luo2022word}  &\checkmark&\checkmark&\checkmark&\checkmark&\checkmark&\checkmark&\checkmark&&\checkmark&&\checkmark&\checkmark&&&\checkmark&\checkmark&& & 80/20 & 12\\
  FLARE22~\cite{ma2023unleashing}  &\checkmark&\checkmark&\checkmark&\checkmark&\checkmark&\checkmark&\checkmark&\checkmark&&\checkmark&\checkmark&\checkmark&\checkmark&\checkmark&&&&& 40/10 & 13\\
  AMOS22~\cite{ji2022amos}  &\checkmark&\checkmark&\checkmark&\checkmark&\checkmark&\checkmark&\checkmark&\checkmark&\checkmark&\checkmark&\checkmark&\checkmark&\checkmark&\checkmark&&&&& 192/48 & 14\\
  \midrule
  \textbf{Sum} & \multicolumn{18}{c|}{} & \textbf{1919/482} & \textbf{18} \\
  \bottomrule
\end{tabular}}}
\label{dataset_details}
\end{table*}

\begin{table*}[t]
\centering
\caption{Available links of selected datasets.}
\setlength{\tabcolsep}{0.9mm}{
\resizebox{0.63\linewidth}{!}{
\begin{tabular}{c|l}
\toprule
 Dataset & \multicolumn{1}{c}{Link} \\ \midrule
 \multirow{3}{*}{M3D-Seg~\cite{bai2024m3d}} & \href{https://github.com/BAAIDCAI/M3D/}{https://github.com/BAAIDCAI/M3D/}\\
 & \href{https://huggingface.co/datasets/GoodBaiBai88/M3D-Seg/}{https://huggingface.co/datasets/GoodBaiBai88/M3D-Seg/} \\
 & \href{https://www.modelscope.cn/datasets/GoodBaiBai88/M3D-Seg/}{https://www.modelscope.cn/datasets/GoodBaiBai88/M3D-Seg/} \\
 \midrule
  MSD-Spleen~\cite{simpson2019large} & \href{http://medicaldecathlon.com/}{http://medicaldecathlon.com/} \\
  Pancreas-CT~\cite{clark2013cancer}  & \href{https://wiki.cancerimagingarchive.net/display/public/pancreas-ct/}{https://wiki.cancerimagingarchive.net/display/public/pancreas-ct/} \\
  LUNA16~\cite{setio2017validation}  & \href{https://luna16.grand-challenge.org/Data/}{https://luna16.grand-challenge.org/Data/} \\ 
  AbdomenCT-1k~\cite{ma2021abdomenct} & \href{https://github.com/JunMa11/AbdomenCT-1K/}{https://github.com/JunMa11/AbdomenCT-1K/} \\
   WORD~\cite{luo2022word}  & \href{https://paperswithcode.com/dataset/word/}{https://paperswithcode.com/dataset/word/} \\
  FLARE22~\cite{ma2023unleashing}  & \href{https://flare22.grand-challenge.org/}{https://flare22.grand-challenge.org/} \\
  AMOS22~\cite{ji2022amos}  & \href{https://amos22.grand-challenge.org/}{https://amos22.grand-challenge.org/} \\
  \bottomrule
\end{tabular}}}
\label{dataset_available}
\end{table*}

\section{Datasets}
\label{appendix_datasets}
\subsection{Statistical Details}
We utilize seven sub-datasets from the M3D-Seg~\cite{bai2024m3d} dataset, a joint cross-modal segmentation dataset with annotated visual targets and textual descriptions, with one to fourteen organs to be segmented. In total, the seven selected sub-datasets have 1919 training samples and 482 testing samples, and these datasets have 18 labels of organs to be segmented: liver (L), spleen (SP), stomach (ST), gallbladder (G), esophagus (E), pancreas (P), duodenum (D), aorta (A), bladder (B), inferior vena cava (IV), left kidney (LK), right kidney (RK), left adrenal gland (LA), right adrenal gland (RA), left femur (LF), right femur (RF), left lung (LL) and right lung (RL). The details of samples  and the target labels of all datasets are as follows:
\begin{enumerate}
    \item The MSD-Spleen~\cite{simpson2019large} consists of 32 training samples and 9 test samples of 1 organ. All slices maintain a consistent dimension of 512$\times$512 pixels, while the z-axis dimension ranges from 31 to 168, with a median of 90.
    \item The Pancreas-CT~\cite{clark2013cancer} consists of 65 training samples and 17 test samples of 1 organ. All slices maintain a consistent dimension of 512$\times$512 pixels, while the z-axis dimension ranges from 181 to 466, with a median of 218. We exclude the pancreatic tumor label in this dataset.
    \item The LUNA16~\cite{setio2017validation} consists of 710 training samples and 178 test samples of 2 organs. All slices maintain a consistent dimension of 512$\times$512 pixels, while the z-axis dimension ranges from 95 to 764, with a median of 237. We exclude the trachea organ and lung nodule labels in this dataset. 
    \item The AbdomenCT-1k~\cite{ma2021abdomenct} consists of 800 training samples and 200 test samples of 5 organs. All slices maintain a consistent dimension of 512$\times$512 pixels, while the z-axis dimension ranges from 31 to 1026, with a median of 103. We separate the original kidney label into left kidney and right kidney labels.
    \item The WORD~\cite{luo2022word} consists of 80 training samples and 20 test samples of 12 organs. All slices maintain a consistent dimension of 512$\times$512 pixels, while the z-axis dimension ranges from 151 to 343, with a median of 202. We exclude the adrenal, colon, intestine, and rectum  organ labels in this dataset.
    \item The FLARE22~\cite{ma2023unleashing} consists of 40 training samples and 10 test samples of 13 organs. All slices maintain a consistent dimension of 512$\times$512 pixels, while the z-axis dimension ranges from 71 to 113, with a median of 106.
    \item The AMOS22~\cite{ji2022amos} consists of 192 training samples and 48 test samples of 14 organs. All slices maintain a consistent dimension of 512$\times$512 pixels, while the z-axis dimension ranges from 64 to 512, with a median of 100. We exclude the prostate organ label in this dataset and we rename the postcava label to vena cava label because they are actually the same organ.
\end{enumerate}

For visual inputs, all the sub-datasets have already been preprocessed in the M3D-Seg dataset, particularly the format of mask files has been harmonized into a sparse matrix storage format and the data has been normalized. Also, all sub-datasets have been split into training and testing sets with a ratio of 8:2. Thus, we follow the original splitting of the joint dataset for the images and mask labels. For the textual information, the joint dataset provides descriptive texts for each organ, including relative and absolute position, shape, size, function, etc., which was generated by external GPT-4o~\cite{hurst2024gpt} and saved in persistent documents for selection. Additionally, we slightly expanded the description sets and generated more diverse descriptions using GPT-4o for some organs. In our experiments, we randomly select two to six sentences with complete and legitimate semantics as textual input. The number of sentences is based on the performance of the segmentation prediction; the worse the performance, the more sentences we add to assist the segmentation. It is also worth mentioning that we build point and bbox prompts for prompt-required models, which are from the ground truths. Firstly, we identify the leftmost, rightmost, topmost, and bottommost positions of the mask to obtain the maximum lengths $\mathit{x}$ and $\mathit{y}$ on the x-axis and y-axis respectively. Then, we randomly select point $(i_0,j_0)$ within the mask, ensuring that points $(i_0+0.1*\mathit{x},j_0)$, $(i_0-0.1*\mathit{x},j_0)$, $(i_0,j_0+0.1*\mathit{y})$ and $(i_0,j_0-0.1*\mathit{y})$ are still within the scope of the mask, which avoids misleading the model if the selected point is located in the margin of the mask. For the bbox prompt, we frame the square composed of four coordinates $(i_1-t_1*\mathit{x},j_1-t_2*\mathit{y})$, $(i_1-t_1*\mathit{x},j_1+t_2*\mathit{y})$, $(i_1+t_1*\mathit{x},j_1-t_2*\mathit{y})$ and $(i_1+t_1*\mathit{x},j_1+t_2*\mathit{y})$, where $(i_1,j_1)$ is the center point of the mask and $t_1,t_2\in[0.1,0.3]$ are random coefficients. It guarantees that the generated bbox could cover the whole organ to be segmented.

To provide more intuitive statistical details, we use the annular chart and bar chart to display the dataset composition of the training and testing samples and the amount and composition of samples of each organ respectively, which are shown in Figure~\ref{dataset_png}. Besides, the information of labels included in each dataset and the available links of datasets are listed in Table~\ref{dataset_details} and Table~\ref{dataset_available} for convenient search.

\subsection{Representative Samples}
\begin{figure*}[t]
    \centering
    \includegraphics[width=0.90\linewidth]{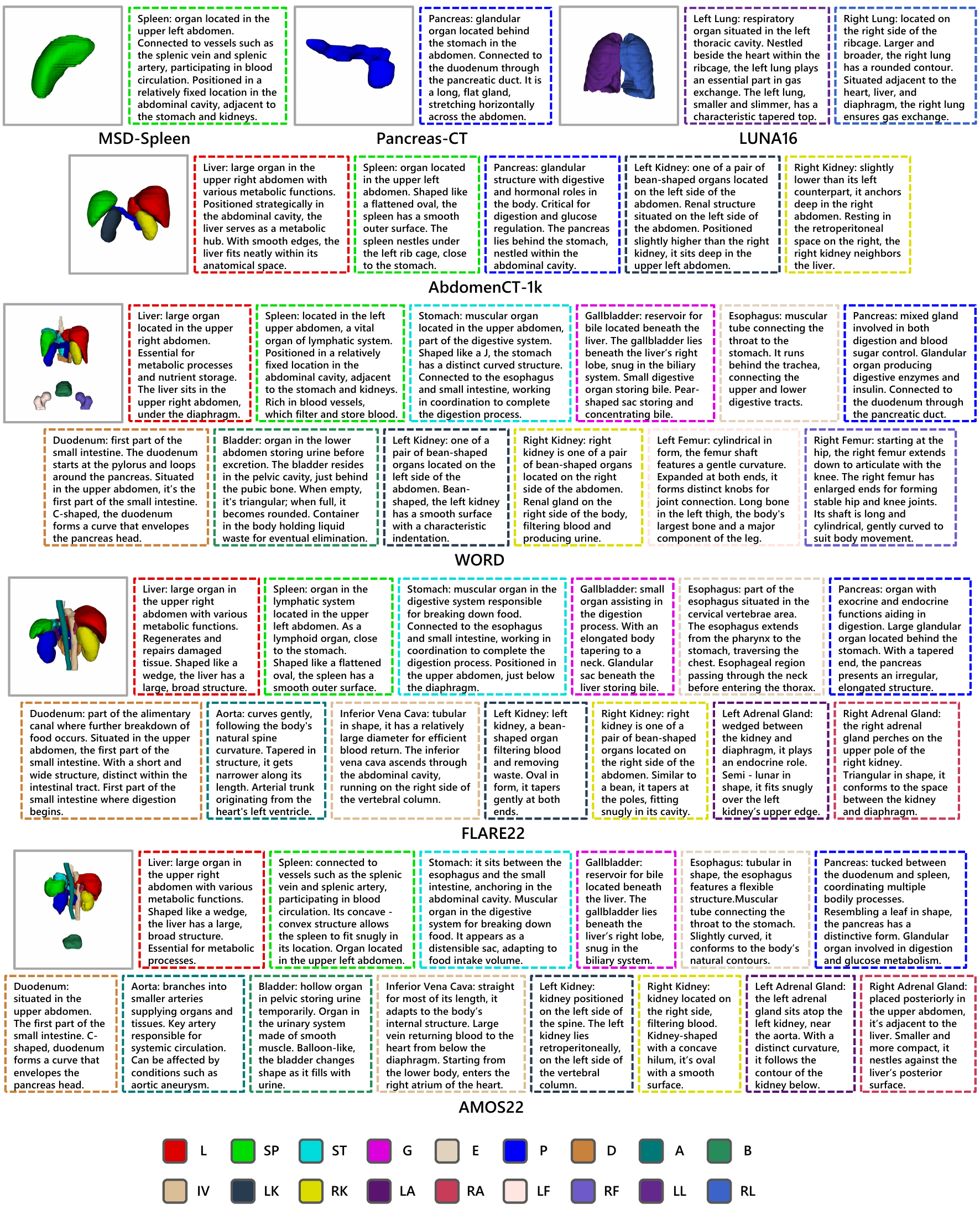}
   \caption{Illustrations of representative samples of seven selected sub-datasets in the joint datasets. Each sample includes a 3D imaging and corresponding descriptive texts for each organ to be segmented.}
    \label{sample}
\end{figure*}

To provide more intuitive demonstrations of the selected cross-modal datasets, we select one representative sample from each dataset, as shown in Figure~\ref{sample}. Each sample is composed of labels of organs to be segmented in 3D view and texts for each organ in the dashed boxes. The color of each organ and the color of each dashed box are one-to-one correspondence.


\section{Experimental Settings}
\subsection{Methodology Details}
\label{methodology_appendix}
\subsubsection*{\textbf{Multi-Head Cross-Attention Mechanism.}} To fuse the features from two modalities or levels, we utilize modified multi-head cross-attention operations for Eq.~\ref{cross_attention1}, \ref{cross_attention2}, \ref{cross_attention3}, which are within the Cross-Modal Semantic Interaction and Semantic Prompt Projector modules respectively. Here, we elaborate the process of depth-wise convolution projection~\cite{lee2022mpvit} and cross-attention operation. Given the first and second parameters $\mathit{X}$ and $\mathit{Y}$, we first perform depth-wise convolutions on $\mathit{Y}$ to capture more localized information. Then, three linear layers further project them into query ($\mathit{Q}$), key ($\mathit{K}$), and value ($\mathit{V}$), respectively. The process is presented as:
\begin{equation}
    \left\{\mathit{Q}, \mathit{K}, \mathit{V}\right\} = \left\{\mathit{W^Q}\mathit{X},\mathit{W^K}DConv_{3}\left(\mathit{Y}\right), \mathit{W^V}DConv_{3}\left(\mathit{Y}\right)\right\},
\end{equation}
where $DConv_{3}\left(\cdot\right)$ denotes $3\times3$ depth-wise convolutions and learnable weight matrices $\{\mathit{W^Q},\mathit{W^K},\mathit{W^V}\}$ are calculated for linear projection. Finally, the cross-attention process could be formulated as:
\begin{equation}
    CrossAttn\left(\mathit{X},\mathit{Y}\right) = Softmax\left(\frac{\mathit{Q}\mathit{K}^{\top}}{\sqrt{d}}\right)\mathit{V},
\end{equation}
where $Softmax\left(\cdot\right)$ represents softmax operation and $\frac{1}{\sqrt{d}}$ denotes the scaling factor. Additionally, we utilize the multi-head mechanism on our cross-attention to extract diverse and rich dependencies between two modalities, and the number of heads is set to 8 in our model.

\subsubsection*{\textbf{Similarity-Sorting Self-Updating Strategy.}} Due to the specificity of 3D medical images, specifically, the spatially intermediate areas contain more radiographic features while the edge areas of the imaging are mainly black-banded, we utilize a similarity-sorting self-updating strategy to ensure high-quality embeddings stored in the memory bank, thus guiding subsequent slices. Firstly, we calculate the similarity score of each slice in a 3D imaging input, which could be formulated as:
\begin{equation}
\mathit{score}_{i} = {\textstyle \sum_{\substack{j=1\\j\ne i}}^{k} sim\left(slice_i, slice_j \right)},
\label{sim_score}
\end{equation}
where $k$ is the count of slices and $sim\left(\cdot,\cdot\right)$ denotes cosine similarity function. Then, we sort the slices by the score as the input order, which means the slice with the higher score will be segmented earlier relatively.

For the self-updating strategy to replace the previous embedding in the memory bank, we first filter the embedding to be updated by the IoU score, which is calculated by the mask decoder alongside with the predicted masks. The embeddings with an IoU score that is lower than the threshold $thresh=0.6$ will be directly discarded because of the less favorable prediction confidence. Then, we calculate a similarity score for the current embedding along with the embeddings stored in the memory bank, the operation is same as in Eq.~\ref{sim_score}. Here, the $k$ is equal to the count of the embeddings stored in the memory bank plus one. It should be noted that if the current embedding has the lowest score, it will not be stored in the memory bank and not replace previous embeddings.

\subsection{Implementation Details}
\label{implementation_appendix}
Following the data augmentation in previous studies~\cite{zhao2024guidednet,guo2024towards,huang2025cat}, random shift and rotation, gamma scaling, and brightness adjustment operations are applied to expand training samples and balance the numbers of each class of organ. We choose Hiera-B+ as the backbone and for the text encoder and image encoder in the Cross-Modal Semantic Interaction module, we adopt ViT-B/16 and CLIP-ViT-B/16 respectively, and fix their parameters. The input imaging is resized to $480\times480$ for the image encoder in this module. As listed in Table~\ref{implement_details} all experiments are conducted on 8 NVIDIA A100-SXM4-80G GPUs in the PyTorch~\cite{NEURIPS2019_bdbca288} framework with AdamW~\cite{loshchilov2017decoupled} optimizer. To make the best use of the global information of visual inputs under the limitations of the memory of GPUs, the input patch size and batch size are set to $(96\times96\times96)$ and 16, with 2 batches per GPU.

The process of training of our CRISP-SAM2 has a total of 400 epochs with two stages: 1) Stage-I: we mainly train the modules except the components of SAM2 to independently optimize them, including the CS-injector, the Cross-Modal Semantic Interaction module (the text and image encoder are frozen), the Semantic Prompt Projector and Local Refiner. This stage has 120 epochs with a warm-up of 80 epochs, and the initial learning rate $1e^{-4}$ with a polynomial decay power of $0.9$, which initializes these components steadily. 2) Stage-II: we further add other components of our model into optimization for 280 epochs, except the four pretrained transformer blocks in the image encoder, and the pretrained text and image encoders in the Cross-Modal Semantic Interaction module. In this stage, the initial learning rate is $2e^{-4}$ to get superior global optimal solutions.

\begin{table}[t]
\centering
\caption{Values of experimental implementations and configurations.}
\setlength{\tabcolsep}{0.9mm}{
\resizebox{0.85\linewidth}{!}{
\begin{tabular}{l|c}
\toprule
 \multicolumn{1}{c|}{Implementation} & Value \\ \midrule
  Initial Values of $\alpha,\beta,\gamma$ & 0.6,0.2,0.2 \\
  Optimizer & AdamW~\cite{loshchilov2017decoupled} \\
  Initial Learning Rate & -\\
  \;\;\;--- Stage-I & $1e^{-4}$\\ 
  \;\;\;--- Stage-II & $2e^{-4}$\\
  Polynomial Decay Power & 0.9\\
  Training Epochs & 400 \\
  \;\;\;--- Stage-I & 120 \\
  \;\;\;\;\;\;--- Warm-up & 80 \\
  \;\;\;--- Stage-II & 280 \\
  GPU & NVIDIA A100-SXM4-80G \\
  GPU Count & 8 \\
  Input Patch Size & $(96\times96\times96)$ \\
  Batch Size & 16 \\
  Batch Size Per GPU & 2 \\
  \bottomrule
\end{tabular}}}
\label{implement_details}
\end{table}

\section{Evaluation Metrics}
All experiments in this work are evaluated by DSC and NSD metrics. The values of these metrics range from 0 to 1; the closer the value is to 1, the better the segmentation performance is. The DSC value is a useful evaluation of whether the segmentation results accurately cover the region of the organ by measuring the overall region overlap. While the NSD focuses on the distance between the surfaces of predicted segmentation and the ground truth. It evaluates the accuracy of segmentation by calculating the average distance between the surfaces and is more sensitive to the accuracy of the segmentation boundaries. The DSC value could be formulated as:
\begin{equation}
\mathit{DSC} = \frac{2 \lvert \mathit{X} \cap \mathit{Y}\,\rvert  }{\lvert\mathit{X}\,\rvert+\lvert\mathit{Y}\,\rvert},
\end{equation}
where $\mathit{X}$ and $\mathit{Y}$ are the predicted mask and ground truth respectively, and $\lvert \mathit{A}\rvert$ represents the cardinality of set $\mathit{A}$. For details of NSD metric, we refer readers to Section~4.6 in \cite{nikolov2018deep}. And the NSD Tolerance is set to 5.

\section{Experimental Results}
\subsection{Comparison Results}
\label{main_experiment_appendix}
\begin{figure*}[t]
    \centering
    \includegraphics[width=0.77\linewidth]{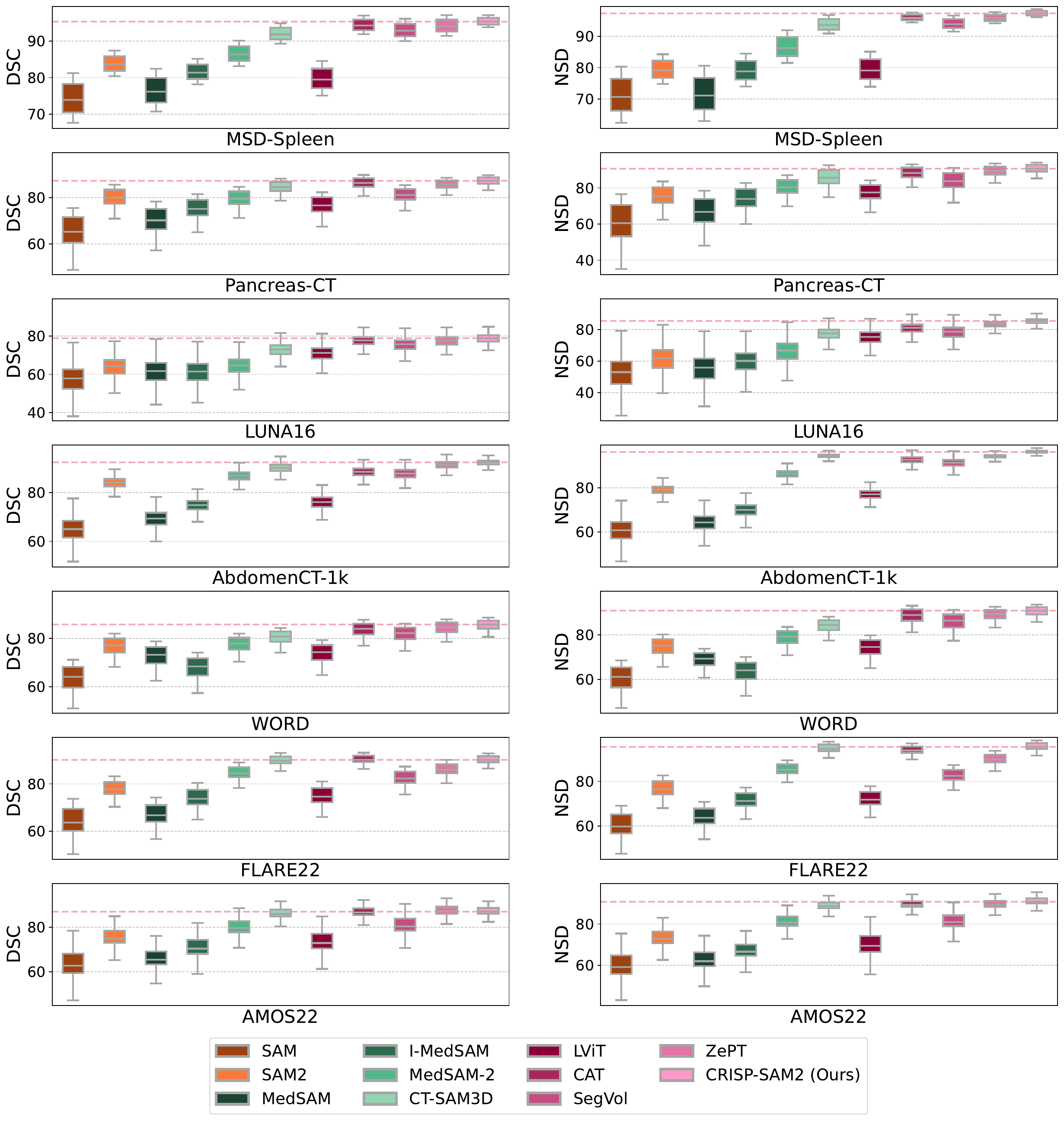}
   \caption{Box plots for comparing experiments of our CRISP-SAM2 and other SOTA methods including DSC and NSD metrics. For intuitive comparison, a dashed line is added at the median of our method.}
    \label{boxplot}
\end{figure*}

\begin{figure*}[t]
    \centering
    \includegraphics[width=0.61\linewidth]{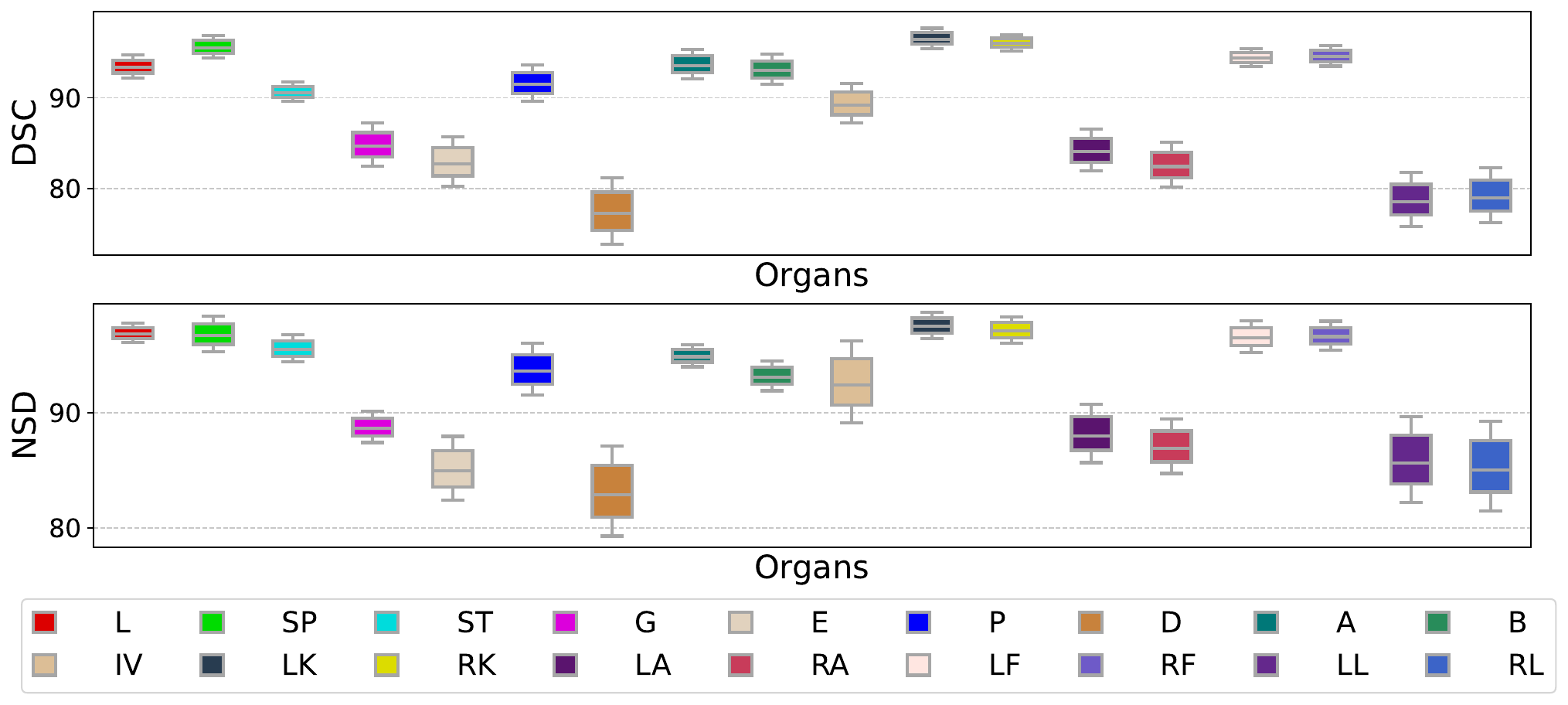}
   \caption{Box plot for the experimental results on DSC and NSD metrics, taking the average values of each organ on the seven selected datasets. }
    \label{organ_boxplot}
\end{figure*}

\begin{figure*}[t]
    \centering
    \includegraphics[width=0.90\linewidth]{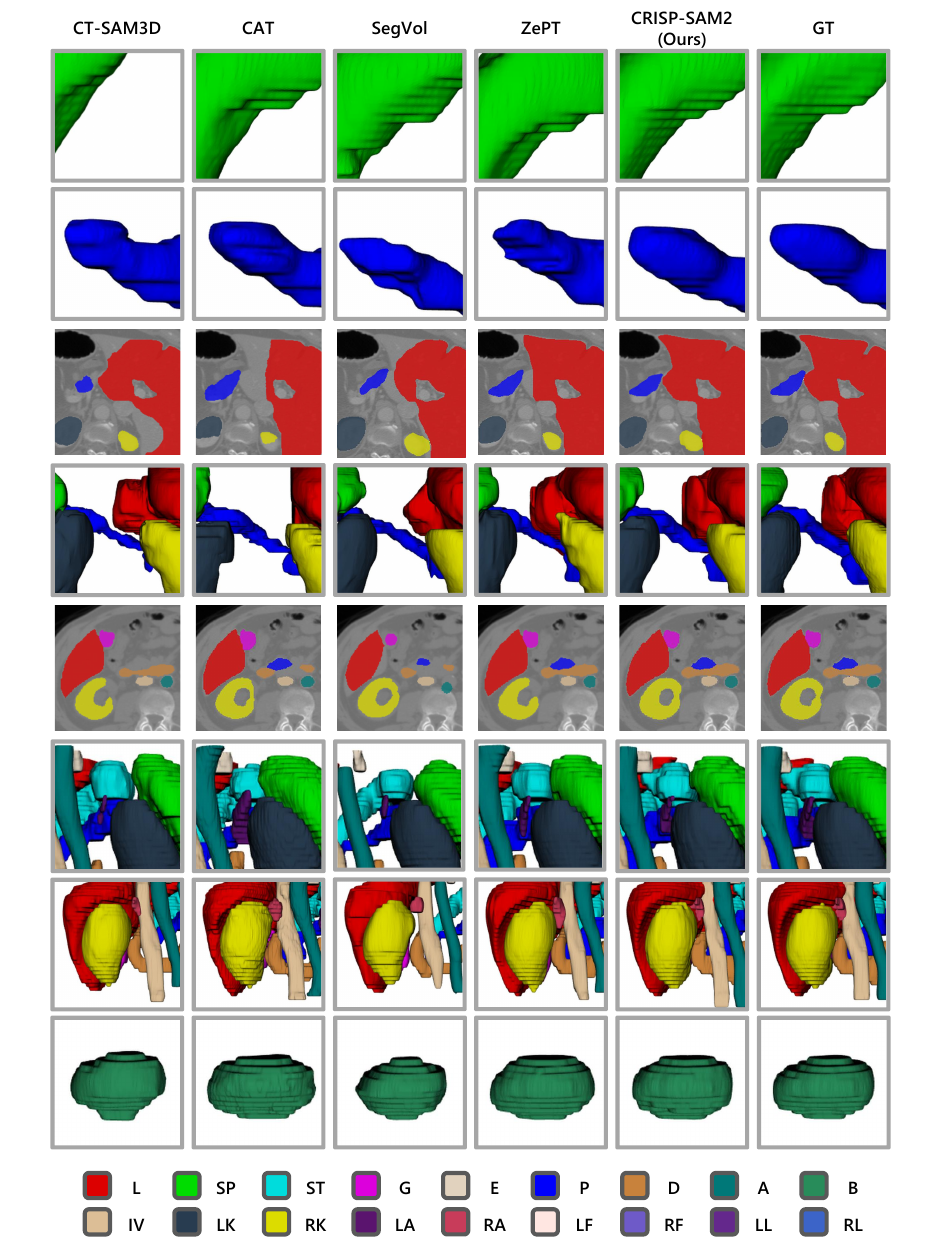}
   \caption{Visualization details of CT-SAM3D~\cite{guo2024towards}, CAT~\cite{huang2025cat}, SegVol~\cite{du2025segvol}, ZePT~\cite{jiang2024zept} and our CRISP-SAM2 on MSD-Spleen~\cite{simpson2019large}, Pancreas-CT~\cite{clark2013cancer}, AbdomenCT-1k~\cite{ma2021abdomenct} and AMOS22~\cite{ji2022amos} datasets.}
    \label{visualization_appendix}
\end{figure*}

In addition to the experimental results elaborated in Section~\ref{comparison_experiments}, we employ box plots comparing the DSC and NSD metrics on each dataset to provide complementary evidence about the performances of other SOTA methods and our CRISP-SAM2. As illustrated in Figure~\ref{boxplot}, our CRISP-SAM2 significantly outperforms other counterparts. In summary, the medians of our method, which is denoted by the dashed lines in each plot, exceed those of other methods in twelve of these fourteen charts. Besides, the size of the rectangle of the box plot embodies the degree of disaggregation of the prediction metrics and segmentation stability of the models. Our CRISP-SAM2 has the smallest rectangles in all plots, demonstrating stable performance on various samples and better robustness. Together with the average values listed in Table~\ref{comparison}, the box plots verify our superior performance in terms of the metrics of segmentation effectiveness. To offer more intuitive evidence, we visualize more detailed predicted segmentation results of CT-SAM3D~\cite{guo2024towards}, CAT~\cite{huang2025cat}, SegVol~\cite{du2025segvol}, ZePT~\cite{jiang2024zept} and our CRISP-SAM2. And we provide the box plots in Figure~\ref{organ_boxplot} for the average values of each organ on the seven selected datasets, showing segmentation performance differences between organs because of their unique characteristics. As depicted in Figure~\ref{visualization_appendix}, the detailed visualization zoom-in intuitively exhibited that our segmentation results are more closely aligned to the ground truths, especially the more accurate local details and boundaries.

\begin{table}[t]
\centering
\caption{Experimental results of unseen labels based on SOTA counterparts and our method.}
\setlength{\tabcolsep}{0.9mm}{
\resizebox{0.92\linewidth}{!}{
\begin{tabular}{c|cc|cc|cc}
\toprule
 \multirow{2}{*}{Method}  & \multicolumn{2}{c|}{Unseen 1} & \multicolumn{2}{c|}{Unseen 2} & \multicolumn{2}{c}{Unseen 3}  \\ \cline{2-3} \cline{4-5} \cline{6-7}  
 & DSC & NSD & DSC & NSD & DSC & NSD \\ \midrule
 SAM2~\cite{ravi2024sam} & 75.41 & 74.46 & 75.85 & 74.36 & 69.47 & 67.20 \\ 
 MedSAM-2~\cite{zhu2024medical} & 77.87 & 78.11 & 76.41 & 76.99 & 69.11 & 68.37 \\
 CT-SAM3D~\cite{guo2024towards} & 73.30 & 75.63 & 74.12 & 76.48 & 71.28 & 72.33 \\ 
 SegVol~\cite{du2025segvol} & 77.70 & 79.34 & \textbf{81.87} & \textbf{83.81} &\textbf{ 73.93} & \textbf{75.32} \\\midrule
 CAT~\cite{huang2025cat} & 76.60 & 77.09 & 77.06 & 79.84 & 70.32 & 72.29 \\
 ZePT~\cite{jiang2024zept} & 77.16 & 78.51 & 79.64 & 81.72 & 70.94 & 73.04\\
\rowcolor{gray!20} 
\textbf{CRISP-SAM2} & \textbf{78.12} & \textbf{80.38} & 80.70 & 83.55 & 72.51 & 75.19  \\ \bottomrule
\end{tabular}}}
\label{zero_shot}
\end{table}

\subsection{Zero-Shot Generalization Performance}
Since the powerful zero-shot generalization ability of SAM2, which could be further enhanced by geometric prompts, we conduct experiments to compare its zero-shot generalization performance with our model. We choose colon, intestine, and rectum in WORD~\cite{luo2022word}, which are not included in training, as "Unseen" labels 1 to 3 respectively for evaluation. For models requiring geometric prompts, i.e. SAM2~\cite{ravi2024sam}, MedSAM-2~\cite{zhu2024medical}, CT-SAM3D~\cite{guo2024towards} and SegVol~\cite{du2025segvol}, we provide points and bboxs, following the previous experiments. As listed in Table~\ref{zero_shot}, it could be observed that these geometric-prompted models have smaller performance reduction when segmenting unseen targets than those without geometric prompts, namely CAT~\cite{huang2025cat}, ZePT~\cite{jiang2024zept} and our model. It is mainly because that geometric prompts are more direct and spatially precise compared to our text-based prompting strategy, requiring exploring inherent relationships between vision and text. While CRISP-SAM2 slightly falls behind SegVol by 0.72\% on the average value of DSC, it has a lead of 0.22\% on NSD evenly when facing the three unseen organs. Compared to baseline SAM2, our method surpasses it 3.53\% and 7.70\% on DSC and NSD respectively. These results demonstrate that our CRISP-SAM2, which utilizes cross-modal interaction and semantic prompting strategies, is capable of generalization performance.

\begin{table}[t]
\centering
\caption{Experiments on the influences of length, quality, type and the linguistic consistency of texts.}
\setlength{\tabcolsep}{0.9mm}{
\resizebox{0.92\linewidth}{!}{
\begin{tabular}{c|c|cc|cc}
\toprule
 \multicolumn{2}{c|}{} & \multicolumn{2}{c|}{Abdomen} & \multicolumn{2}{c}{AMOS22}  \\ \cline{3-4} \cline{5-6}
 \multicolumn{2}{c|}{} & DSC & NSD & DSC & NSD\\ \midrule
 \multirow{3}{*}{Length} & Only Label & 88.94 & 91.25 & 81.59 & 85.91 \\ 
 & Short & 91.13 & 95.32 & 85.40 & 89.57 \\ 
 & \cellcolor{gray!20} \textbf{Long (Ours)} & \cellcolor{gray!20} \textbf{92.28} & \cellcolor{gray!20} \textbf{96.40} & \cellcolor{gray!20} \textbf{86.88} & \cellcolor{gray!20} \textbf{90.74} \\ \midrule
 \multirow{3}{*}{Quality} & Irrelevant & 87.27 & 90.60 & 80.56 & 84.47\\
 & Mixed & 91.95 & 95.77 & 85.86 & 89.10 \\
 & \cellcolor{gray!20} \textbf{Relevant (Ours)} & \cellcolor{gray!20} \textbf{92.28} & \cellcolor{gray!20} \textbf{96.40} & \cellcolor{gray!20} \textbf{86.88} & \cellcolor{gray!20} \textbf{90.74} \\ \midrule
 \multirow{5}{*}{Type} & Position & 91.74 & 95.85 & 85.75 & 89.86 \\
 & Shape & 91.02 & 95.87 & 84.91 & 89.40 \\ 
 & Size & 90.79 & 94.82 & 84.73 & 88.87 \\
 & Function & 90.91 & 94.96 & 85.19 & 89.38 \\
 & \cellcolor{gray!20} \textbf{Mixed (Ours)} & \cellcolor{gray!20} \textbf{92.28} & \cellcolor{gray!20} \textbf{96.40} & \cellcolor{gray!20} \textbf{86.88} & \cellcolor{gray!20} \textbf{90.74} \\ \midrule
 Linguistic  & \XSolidBrush & 92.04 & 95.93 & 86.26 & 90.51 \\
 Consistency & \cellcolor{gray!20} \CheckmarkBold \textbf{(Ours)} & \cellcolor{gray!20} \textbf{92.28} & \cellcolor{gray!20} \textbf{96.40} & \cellcolor{gray!20} \textbf{86.88} & \cellcolor{gray!20} \textbf{90.74} \\ \bottomrule
\end{tabular}}}
\label{textual_inputs}
\end{table}

\subsection{Analysis of Textual Inputs}
To better discuss the influence of different textual inputs on segmentation performance of our model, we design comprehensive experiments, assessing the influences of length, quality, type, and linguistic consistency between training and testing data. To evaluate influences of the length of texts, we set three groups of different lengths: only label, including solely the organ names; short, which contains only one sentences; long, following the original design of our model. For assess the impact of quality, we compare the performance of irrelevant descriptions from other organs and mix them with relevant texts in equal parts as another control group. In terms of the types of content, we separate the four main descriptive types, namely relative and absolute position, shape, size and function, to appraise contributions of specific types. Also, we conduct experiments on inconsistent textual inputs between training and testing, and the texts of control group are generated by models Llama-3~\cite{grattafiori2024llama} other than GPT-4o~\cite{hurst2024gpt}.

\begin{figure}[t]
    \centering
    \includegraphics[width=0.93\linewidth]{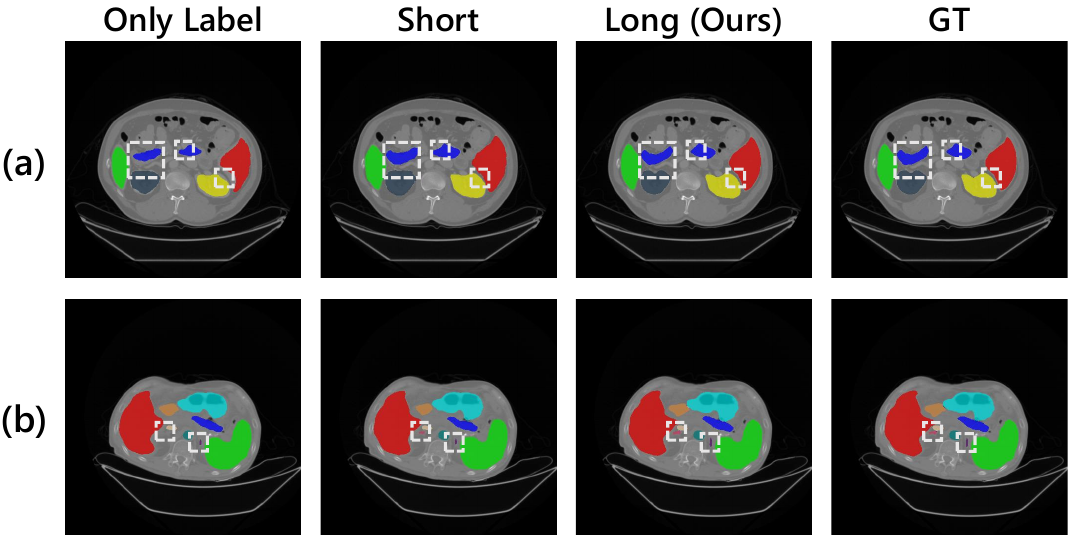}
   \caption{Visualizations of texts of various lengths on (a) AbdomenCT-1k~\cite{ma2021abdomenct} and (b) AMOS22~\cite{ji2022amos} datasets, demonstrating the impacts of textual information.}
    \label{qualitative}
\end{figure}

The experimental results of these factors are demonstrated in Table~\ref{textual_inputs}. First, the length of input texts has a noticeable impact on performance with a positive correlation between them, which could also be verified in Figure~\ref{qualitative}. When we limit the textual length to only one sentence, it exhibits a reduction of 1.32\% on DSC and 1.13\% on NSD, and this decrease further expands to 4.32\% and 4.99\% respectively when the input is limited to only organ names due to exceptionally scarce semantics. Second, poor quality of texts with irrelevant information also restricts segmentation performance. Inputting completely irrelevant information decreases both DCS and NSD by over 5.5\% and 6.0\%, however, the influence of mixing irrelevant texts with relevant ones is relatively limited, showing excellent robustness. Third, certain types of textual information contribute differently to segmentation. For instance, the positional information benefits the most, while the descriptions of size contribute the least. And including texts about the shape and function of organs brings more improvements on NSD compared to DSC, resulting in more accurate boundaries and details. Notably, incorporating any of these four main textual types offers richer semantics than only providing labels to guide the segmentation, and utilizing the mixture of various textual types as input achieves the best performance. Last, the inconsistency of the textual during training and inference periods causes fairly small reductions on the two metrics, which are 0.43\% and 0.35\% on DSC and NSD respectively, verifying great generalization ability and robustness of our model.

To thoroughly evaluate the impact of textual length, we conduct qualitative experiments using texts of various levels of length. As illustrated in Figure~\ref{qualitative}, the inclusion of more comprehensive textual information enables our model to generate more precise segmentation masks. This is particularly evident in handling irregularly shaped boundaries and small organs, as highlighted in the dashed boxes of panels (a) and (b), emphasizing the significance of textual information in enhancing segmentation accuracy.

\begin{table}[t]
\centering
\caption{Comparisons with baselines on the computational cost and inference efficiency. FPS: frames per second. Param.: parameters of the model. FLOPs: floating point operations.}
\setlength{\tabcolsep}{0.9mm}{
\resizebox{0.85\linewidth}{!}{
\begin{tabular}{c|c|c|c|c}
\toprule
& \multirow{2}{*}{SAM-L~\cite{kirillov2023segment}} & SAM2~\cite{ravi2024sam} & SAM2~\cite{ravi2024sam} & CRISP-SAM2\\ 
& & (Hiera-B+) & (Hiera-L) & (Hiera-B+) \\ \midrule
\multirow{2}{*}{Param.} & \multirow{2}{*}{312M} & \multirow{2}{*}{81M} & \multirow{2}{*}{224M} & 81M + 196M \\
& & & & = 277M \\ 
FLOPs & 2690 & 560 & 1490 & 1620 \\
FPS & 4.2 & 36.5 & 24.1 & 20.9 \\ 
\bottomrule
\end{tabular}}}
\label{computational_cost}
\end{table}

\subsection{Computational Cost and Efficiency}
In order to obtain more intuitive comparisons, we compare our CRISP-SAM2 with three baselines: SAM-L~\cite{kirillov2023segment} and SAM2 with Hiera-L backbones~\cite{ravi2024sam}, which are selected in comparative experiments, and SAM2 with Hiera-B+ backbones, which is much more lightweight. As illustrated in Table~\ref{computational_cost}, our model has 23.7\% more parameters than SAM2 using Hiera-L, because of the incorporation of extra modules, e.g. cross-modal semantic interaction, whose additional parameters are mainly from the pretrained text encoder and image encoder. Compared to the encoders and the Hiera architecture, the cross-attention layers or other operations bring limited parameters, which are less than 15\% in total. However, the FLOPs only increases 8.7\% to 1620G, enabling to train our model on the same GPU limitations with SAM2 (Hiera-L). Besides, the speed during inference decreases 13.3\%, remaining high inference efficiency, especially compared with SAM based methods.

\section{Limitations}
Although our proposed CRISP-SAM2 has achieved SOTA performance on various datasets, there is still limitations to be solved: (1) Relatively limited data: although we conduct experiments on seven public datasets, our total training and testing samples are still insufficient compared to works like SegVol~\cite{du2025segvol}. Besides, the richness and numbers of sentences in our textual descriptions could be further expanded. (2) Not outstanding enough zero-shot generalization: although our model still achieve superior performance of segmenting unseen targets, the leading advantage has been reduced, compared to supervised segmentation. (3) Relatively high computational cost and low inference speed: our model has slightly larger parameters, and vaguely slower speed during inference. (4) Failures when textual descriptions mismatching visual inputs: some organs sometimes shift due to lesions or congenital abnormalities as well as when images are flipped. We leave these limitations to be tackled in our future works.

\end{document}